\documentclass[a4paper,fleqn, usenatbib]{mnras}

\usepackage[T1]{fontenc}
\usepackage{ae,aecompl}
\usepackage{longtable}

\usepackage{graphicx}	
\usepackage{amsmath}	
\usepackage{amssymb}	

\title[Dust Reverberation Mapping]{Dust Reverberation Mapping of 
Z229$-$15}

\author[Mandal et al.]{Amit Kumar Mandal$^{1,2}$\thanks{E-mail: amitkumar@iiap.res.in},
Suvendu Rakshit$^{3,4}$,
C. S. Stalin$^{2}$, Dominika Wylezalek$^{5}$,
\newauthor
Markus Kissler Patig$^{5}$,
Ram Sagar$^{2}$,
Blesson Mathew$^{1}$,
S. Muneer$^{2}$,
Indrani Pal$^{2}$
\\\\
$^{1}$Department of Physics, CHRIST (Deemed to be University), Hosur Road, Bangalore 560 029, India\\
$^{2}$Indian Institute of Astrophysics, Block II, Koramangala, Bangalore, 560 034, India\\
$^{3}$Aryabhatta Research Institute of Observational Sciences, Manora Peak, Nainital 263002, India\\
$^{4}$Finnish Centre for Astronomy with ESO (FINCA), University of Turku, Quantum, Vesilinnantie 5, 20014 University of Turku, Finland\\
$^{5}$European Southern Observatory, Karl-Schwarzschildstr. 2, D-85748 Garching bei Mu\"unchen, Germany \\
\\
}

\date{Accepted 08 December 2020. Received 08 December 2020; in original form 14 September 2020}

\pubyear{2020}

\begin{document}
\label{firstpage}
\pagerange{\pageref{firstpage}--\pageref{lastpage}}
\maketitle

\begin{abstract}

We report results of the dust reverberation mapping (DRM) on the 
Seyfert 1 galaxy Z229$-$15 at $z$ = 0.0273. Quasi-simultaneous photometric
observations for a total of 48 epochs were acquired during the period 
2017 July to 2018 December in B, V, J, H and $\mathrm{K_{s}}$ bands.  The  calculated  
spectral index ($\alpha$) between
B and V bands for each epoch was used to correct for
the accretion disk (AD) component present in the infrared light 
curves.  The observed $\alpha$ ranges between $-$0.99 and 1.03. Using cross-correlation function analysis we found significant time
delays between the optical V and the AD corrected J, H and $\mathrm{K_s}$ light curves. 
The lags in the rest frame of the source are  
$12.52^{+10.00}_{-9.55}$ days (between V and J), $15.63^{+5.05}_{-5.11}$ days 
(between V and H) and $20.36^{+5.82}_{-5.68}$ days (between V and $\mathrm{K_{s}}$).
Given the large error bars, these lags are consistent with each other. However,
considering the lag between V and $\mathrm{K_s}$ bands to represent the inner edge of 
the dust torus, the torus in Z229$-$15 lies at a distance of  
0.017 pc from the central ionizing continuum. This is smaller than
that expected from the radius luminosity (R$-$L) relationship known from DRM. 
Using a constant $\alpha$ = 0.1 to account for the AD component, as is 
normally done in DRM,  the deduced radius ( 0.025 pc) lies close to the 
expected R$-$L relation. However, usage of constant $\alpha$ in DRM  is 
disfavoured as the $\alpha$ of the ionizing continuum changes with the flux 
of the source.
       
\end{abstract}




\section{Introduction}

Active Galactic Nuclei (AGN) are amongst the most luminous objects 
(10$^{42}$ $-$ 10$^{48}$ erg s$^{-1}$)
in the Universe and emit energy over all wavelengths. They are
believed to be powered by accretion of matter onto super massive black hole
(SMBH; 10$^{6}$ $-$ 10$^{10}$ M$_{\odot}$) located at the centers of galaxies. 
The process of accretion forms an accretion disk around the SMBH that 
radiates predominantly in the  ultra-violet (UV) and optical wavelengths 
\citep{1964ApJ...140..796S,1969Natur.223..690L,1973A&A....24..337S}.
The broad line region (BLR) that lies outside
the accretion disk, produces  the line emission due to reprocessing of
the  UV/optical radiation from the accretion disk. Further out from
the BLR is the obscuring torus that is responsible for the thermal
infrared emission.  Among the different types of AGN are the Seyfert 
galaxies \citep{1943ApJ....97...28S}. Depending on the presence or absence of 
broad emission lines in their spectra, Seyfert galaxies are divided into 
Seyfert 1 and Seyfert 2 galaxies. According to the Unified model of AGN, the 
obscuring torus located within a few parsecs from 
the central SMBH is responsible for the separation 
of Seyfert galaxies into Seyfert 1 and Seyfert 2 category \citep{1993ARA&A..31..473A,1995PASP..107..803U}. For a typical Seyfert galaxy with a UV luminosity 
of 10$^{42}$ $-$ 10$^{44}$ erg s$^{-1}$, BLR lies at about $\sim$0.01 pc from 
the accretion disk and  the inner edge of the dust torus surrounding the BLR, can extend 
from  0.01 $-$ 0.1 pc. So, the central regions of AGN are very compact and not 
possible to image directly. Therefore, it is difficult to know by direct means the dimension of the BLR and the dust torus in an AGN. 

Two methods are currently available to determine the inner extent of the
dust torus in an AGN. The first one called reverberation mapping
\citep{1982ApJ...255..419B,1993PASP..105..247P} uses the intrinsic 
characteristic of AGN, namely its flux variability. The optical/UV continuum 
for an AGN is known to show flux variations on time sales of days to years
\citep{1995ARA&A..33..163W}. The response of the
infrared K band signal to the optical continuum variability is delayed by a 
time interval $\mathrm{\delta t}$ that characterizes the inner edge of the dust torus
as $R_{\mathrm{torus}} < \mathrm{c \times \delta t}$, where c is the speed of light.
This method of determining the extent of the torus via monitoring observations
is called dust reverberation mapping (DRM). This is an expensive method in terms
of the observations required, however, insensitive to the distance of the 
source. Using DRM through optical and near infrared (NIR) K band observations, 
presently the extent of the torus has been measured for about
40 AGN \citep{2014ApJ...788..159K,2014A&A...561L...8P,2015A&A...576A..73P,2018MNRAS.475.5330M,2018A&A...620A.137R,2019ApJ...886..150M,2020AJ....159..259S}. 
Also, recently using optical and mid infrared (MIR) observations, 
\cite{2019ApJ...886...33L} reported results of torus size for
most of the Palomar-Green quasars. The MIR time lags were found to 
follow the relation $\Delta t \propto L^{0.5}$, and the average torus size 
increased with wavelength. The second method available today to measure the 
extent of the torus in AGN is via NIR  or MIR 
interferometry.  This method
has also been successful in measuring the size of the torus in 
about two dozen AGN 
\citep{2009A&A...507L..57K,2011A&A...527A.121K,2020A&A...635A..92G}. 
This method is suitable to only nearby bright AGN, while DRM can be used to
measure the torus size in any AGN. However, there are systematic 
differences in the torus size measured by these two methods. The size 
measured from interferometric observations are always larger than that found 
from DRM. Also, the dust torus size from NIR
interferometric observations do not follow the relation of 
$R_{\mathrm{torus}} \propto L^{0.5}$ \citep{2014ApJ...788..159K} known from
K band DRM observations, instead the half light radius R$_{1/2}$ which is used 
as a representative size of the torus varies as $R_{1/2} \propto  L^{0.21}$ 
\citep{2011A&A...536A..78K}. Recently, from DRM observation of 22 $z$ $<$
0.6 quasars \cite{2019ApJ...886..150M} found $R_{\mathrm{torus}} \propto L^{0.424}$.

In spite of the differences in the torus size obtained from DRM and
interferometric observations, the strong correlation between the dust lag 
and the optical luminosity suggests that AGN can be used as  a standard candle \citep{1999AstL...25..483O, 2001ASPC..224..149O, 2014ApJ...784L..11Y, 2014ApJ...784L...4H, 2017ApJ...838L..20H}. \citet{2017ApJ...838L..20H} started a large DRM program, 'VEILS' (VISTA 
Extragalactic Infrared Legacy Survey) that will observe about 1350 targets in 
the redshift range of $0.1 < z < 1.2$  to use dust lag as standard candle to 
constrain cosmological parameters. However, this program will eventually miss 
the objects in the local universe, which are important to determine the 
normalization parameter of the AGN distance moduli \citep{2017ApJ...838L..20H}.
As a complement to the VEILS program, in the nearby Universe, we have
started a monitoring project called the the REverberation Mapping of Active
galactic nuclei Program 
(REMAP; \citealt{2018MNRAS.475.5330M,2019BSRSL..88..158M}). For this
program we are using the 2 m Himalayan Chandra Telescope (HCT) at Hanle, India 
to carry out observations on a suitably selected sample of eight sources
taken from the catalogue of \citet{2015PASP..127...67B} that 
has spectroscopic lag measurements of the BLR.
The results on the first target from the REMAP program, namely
H0507+164 was published in \citet{2018MNRAS.475.5330M}. Here, we present the
results of the second source, namely Z229$-$15, a local
Seyfert 1 galaxy with redshift $z$=0.0273.
Located at $\alpha_{2000}$ = 19:05:25.94 and $\delta_{2000}$ = +42:27:39.76, 
Z229$-$15 has a black hole mass of 
$1.00^{+0.19}_{-0.24} \times 10^7 \, M_{\odot}$ 
\citep{2011ApJ...732..121B} and a {\it Gaia} G band brightness of 16.44 mag.  In Section 2, we 
describe the
observation and data reduction processes. The analysis is given in Section 3.
In Section 4 we discuss the results of this work followed by the summary in
Section 5. For the cosmological parameters, we
assumed $\mathrm{H_{0} = 73 \, km \, s^{-1} \, Mpc^{-1}}$, $\mathrm{\Omega_m = 0.27}$, and $\mathrm{\Omega_\lambda = 0.73}$ \citep{2014ApJ...788..159K}.

\section{Observation and data reduction}
The photometric observations in  the optical B and V bands and 
 the infrared J, H, 
and $\mathrm{K_{s}}$ bands were carried out during the period 2017 July to 2018 
December for a total of 48 epochs using the Himalayan Chandra Telescope (HCT). Optical observations
were carried out using the Himalayan Faint Object Spectrograph and 
Camera (HFOSC) mounted at the Cassegrain focus of HCT. The camera has 
a 2048 $\times$ 4096 SiTe CCD chip  with a gain and readout noise of 1.22 electrons/ADU and 4.8 electrons, respectively. Each pixel of 
the CCD covers a region of 0.296$^{\prime\prime}$ in the sky. The observations 
were carried out in binned mode using only the central 
2048 $\times$ 2048 region of the CCD, thus
covering a field of view of 10$^{\prime}$ $\times$ 10$^{\prime}$. The exposure 
time in B and V bands is 150 sec and 50 sec, respectively. The NIR 
observations in  the J, H, and $\mathrm{K_{s}}$ bands were done after the V band observations at each epoch 
using the TIFR Near Infrared Spectrometer (TIRSPEC) mounted on one of the side 
ports of HCT \citep{2014JAI.....350006N}. The detector used in TIRSPEC is a 
1024 $\times$ 1024 HgCdTe array with a pixel size of 18 $\mu$m covering a 
field of view of 5$^{\prime}$ $\times$ 5$^{\prime}$. It has a readout noise and gain of 
21.5 electrons and 5 electrons/ADU, respectively. The NIR images were taken in 
3 dither positions consisting of five exposures each of 20 sec in each of
 the three NIR filters namely J, H, and $\mathrm{K_{s}}$. Sky regions 
were also observed  
in the same dithering pattern as the science frames to generate the master 
sky frame.

\subsection{Data reduction}
The optical data were reduced using \textsc{iraf} (Image Reduction and Analysis Facility)
and \textsc{midas} (Munich Data Analysis System). We followed the standard procedures 
for image reduction, such as bias  subtraction, dark subtraction and flat-fielding. Cosmic 
rays were removed using \textsc{midas}. The NIR images were reduced using 
TIRSPEC NIR Data Reduction Pipeline \citep{2014JAI.....350006N}.  
The pipeline produces  the final combined images after 
 performing dark subtraction and flat fielding on the raw image frames.
\subsection{Optical photometry}
\begin{figure}
	\includegraphics[scale=0.6]{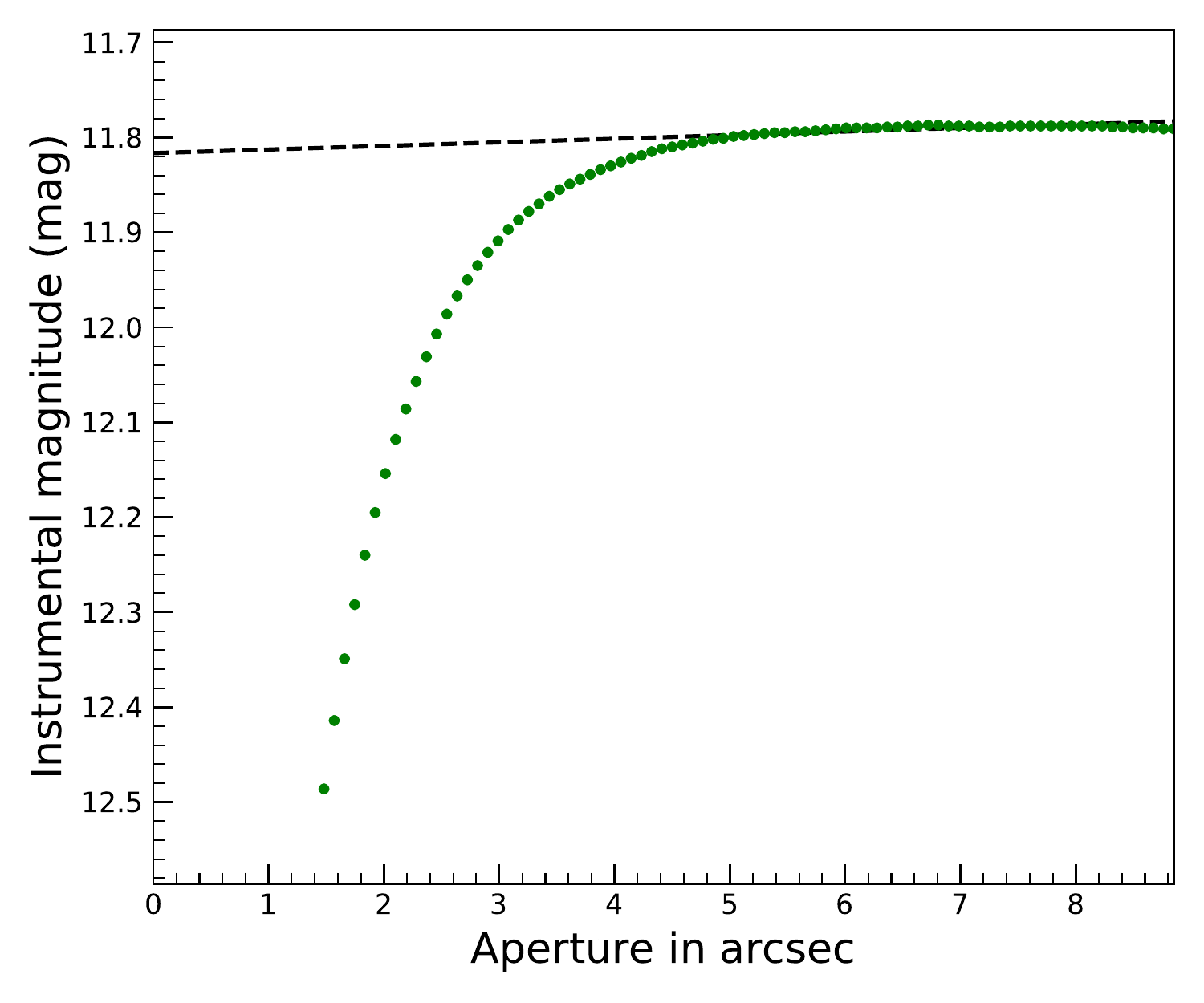} 
	\caption{Growth curve for a  comparison star in V band. The dashed
line is the linear least squares fit to the points with 
aperture sizes between 4 and 6 times the
FWHM.}
\label{fig:fig-1}
\end{figure}

\begin{figure}
	\includegraphics[scale=0.58]{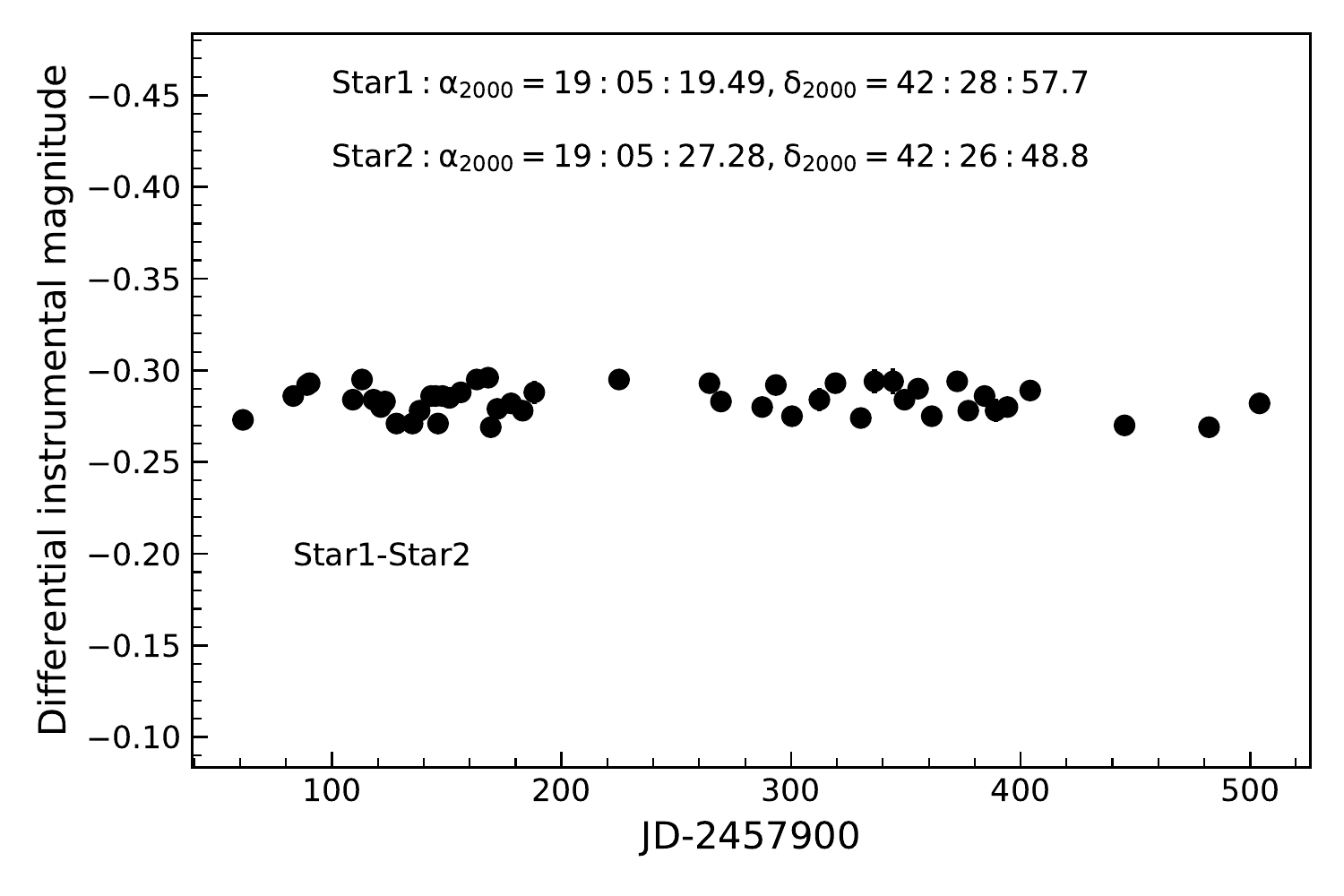} 
	\caption{ Differential light curve of two stars, Star 1 
	($\mathrm{\alpha_{2000} = 19:05:19.49}$, $\mathrm{\delta_{2000} = 42:28:57.7}$) and Star 2 ($\mathrm{\alpha_{2000} = 19:05:27.28}$, $\mathrm{\delta_{2000} = 42:26:48.8}$) in V band present in the observed field of Z229$-$15.}
\label{fig:fig-2}
\end{figure}

The objects in the observed image frames were detected using the 
{\it daofind} task in \textsc{iraf}. Photometry of those detected objects were
then carried out using the {\it phot} task in \textsc{iraf}. Of the detected objects
we selected two comparison stars having similar brightness to the AGN to
carryout out differential photometry and to bring the instrumental magnitude
to the standard system. The instrumental magnitudes were obtained in several 
concentric circular apertures centered on the comparison stars starting
from FWHM to about 8 times the FWHM. Growth curves were generated and the final
instrumental magnitude for each of  the comparison star in an epoch is 
obtained by the curve of growth (COG) method. This adoption of the COG method
ensures that the total flux from a point source is measured. The 
growth curve for a comparison star is shown in  Fig. \ref{fig:fig-1}.
We fit a straight line using the photometric points between 4 and 6 times the
FWHM of the point source (at which point the COG smoothly merges with the
background) and the intercept of that line (shown as  a dashed line
in Fig. \ref{fig:fig-1}) was taken as the magnitude of
the point source.  The two comparison stars that were used to get the correction 
factors to bring the magnitudes of  Z229$-$15 to the standard system are found 
to be steady during the duration of our observations. The differential light
curve  (DLC) of these two stars in the V band are shown in Fig. \ref{fig:fig-2}. The DLC has a  
standard deviation of 0.008 mag, while the mean error
of the photometric points in the DLC is 0.004 mag. Given that the 
photometric errors given by the {\it phot} task in IRAF is 
an underestimate by a factor of about 1.75 \citep{1995MNRAS.274..701G}, 
the standard
deviation of the DLC of the two stars is 
consistent with the photometric error, which confirms that the
two stars are not variable during the period of our observations.

\begin{figure*}
\resizebox{5.8cm}{5.5cm}{\includegraphics{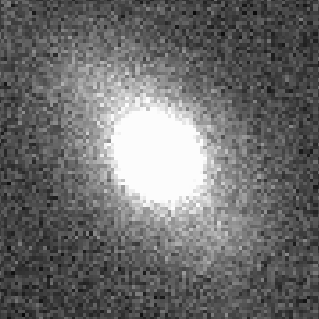}} 
\resizebox{5.8cm}{5.5cm}{\includegraphics{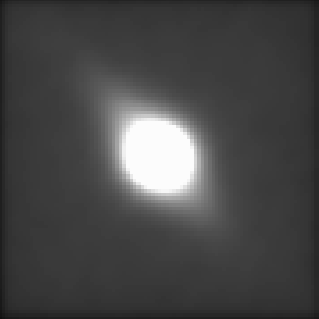}}
\resizebox{5.8cm}{5.5cm}{\includegraphics{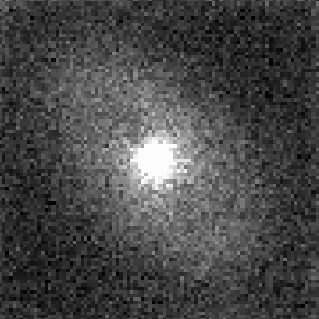}} 
\caption{Observed V band image of Z229$-$15 (left). Modeled galaxy image (middle) and residual image containing the AGN at the center (right) obtained from GALFIT.}
\label{fig:fig-3}
\end{figure*}

\begin{figure}
\resizebox{9cm}{7cm}{\includegraphics{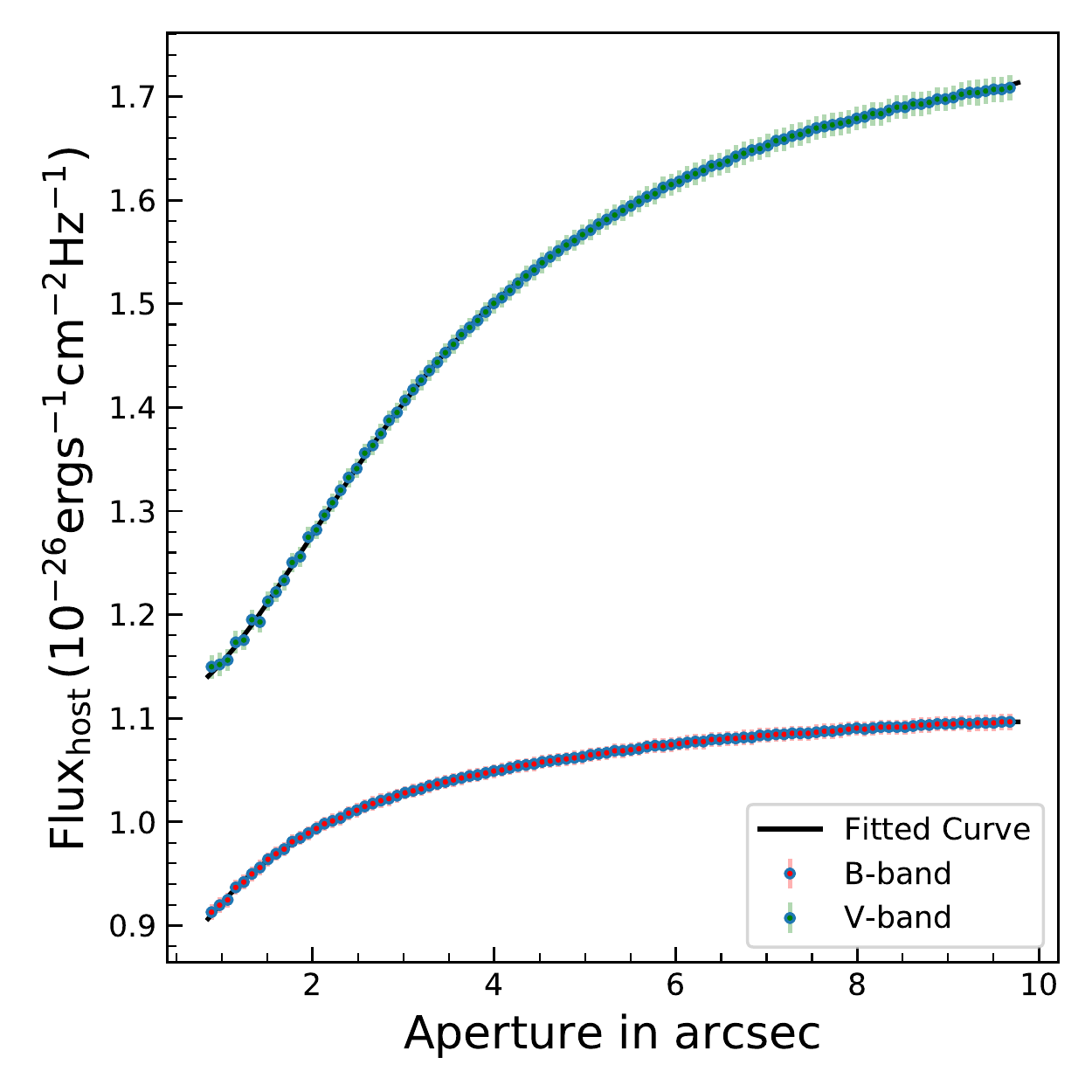}} 
\caption{Host-galaxy flux as a function of aperture size. The points with error bar are the galaxy-flux values and the solid black lines are the best polynomial fits.}
\label{fig:fig-4}
\end{figure}

\begin{figure}
\resizebox{8.5cm}{7cm}{\includegraphics{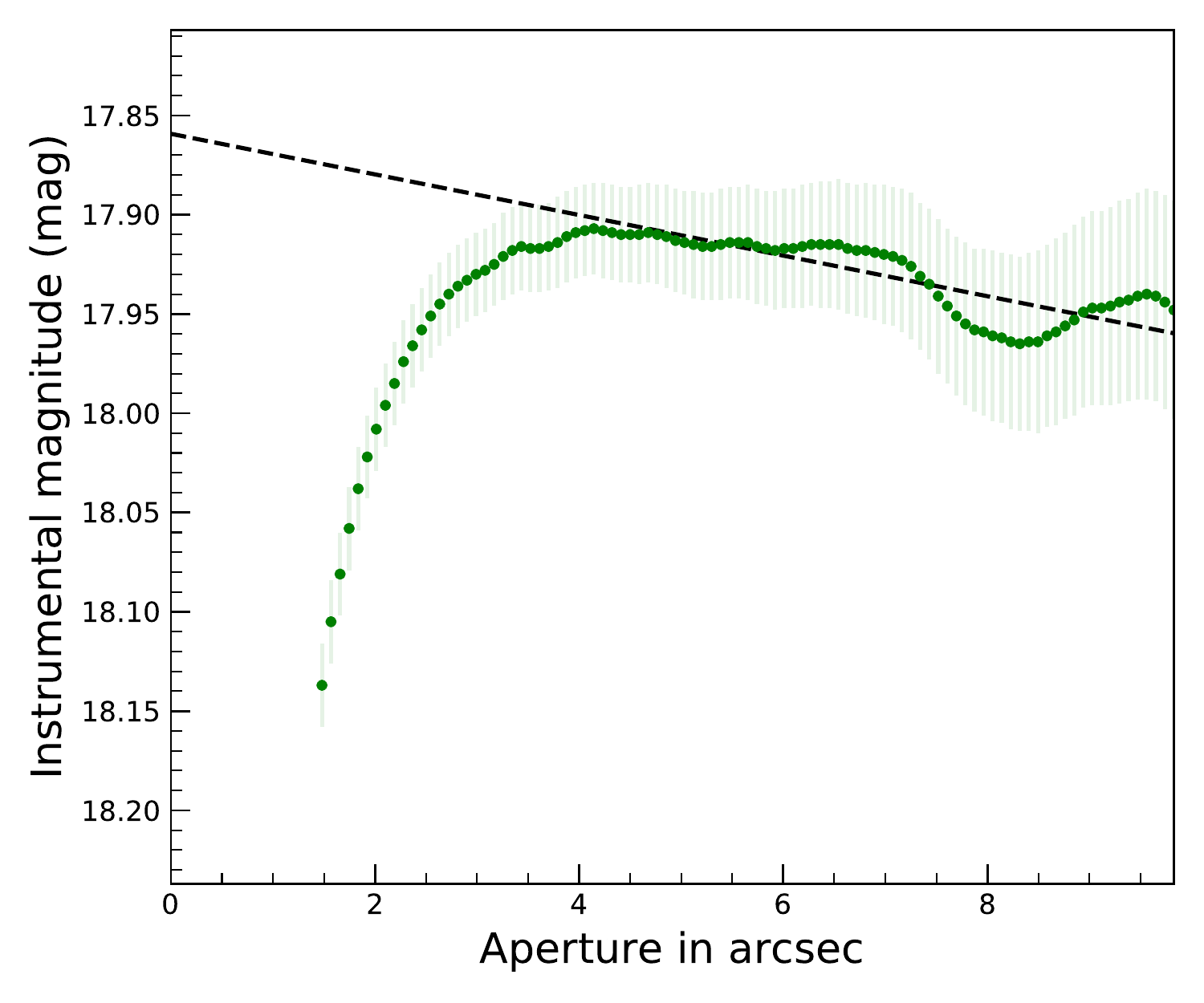}}
	\caption{Growth curve for a comparison star in $\mathrm{K_s}$ band. The dashed line is the linear least squares fit to the points with 
aperture sizes between 4 and 6 times the FWHM.}
\label{fig:fig-5}
\end{figure}

\begin{figure}
\resizebox{9cm}{7cm}{\includegraphics{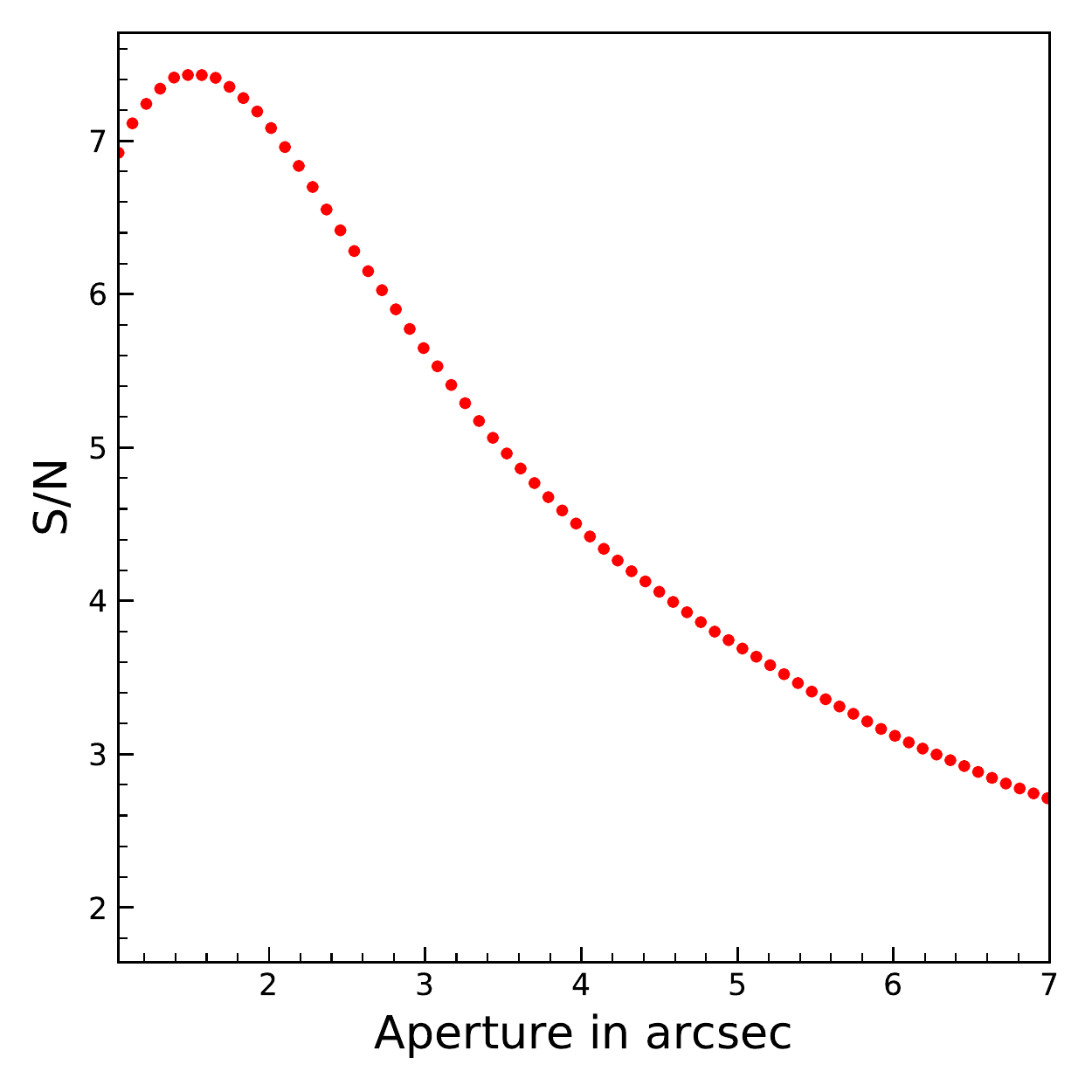}}
\resizebox{9cm}{7cm}{\includegraphics{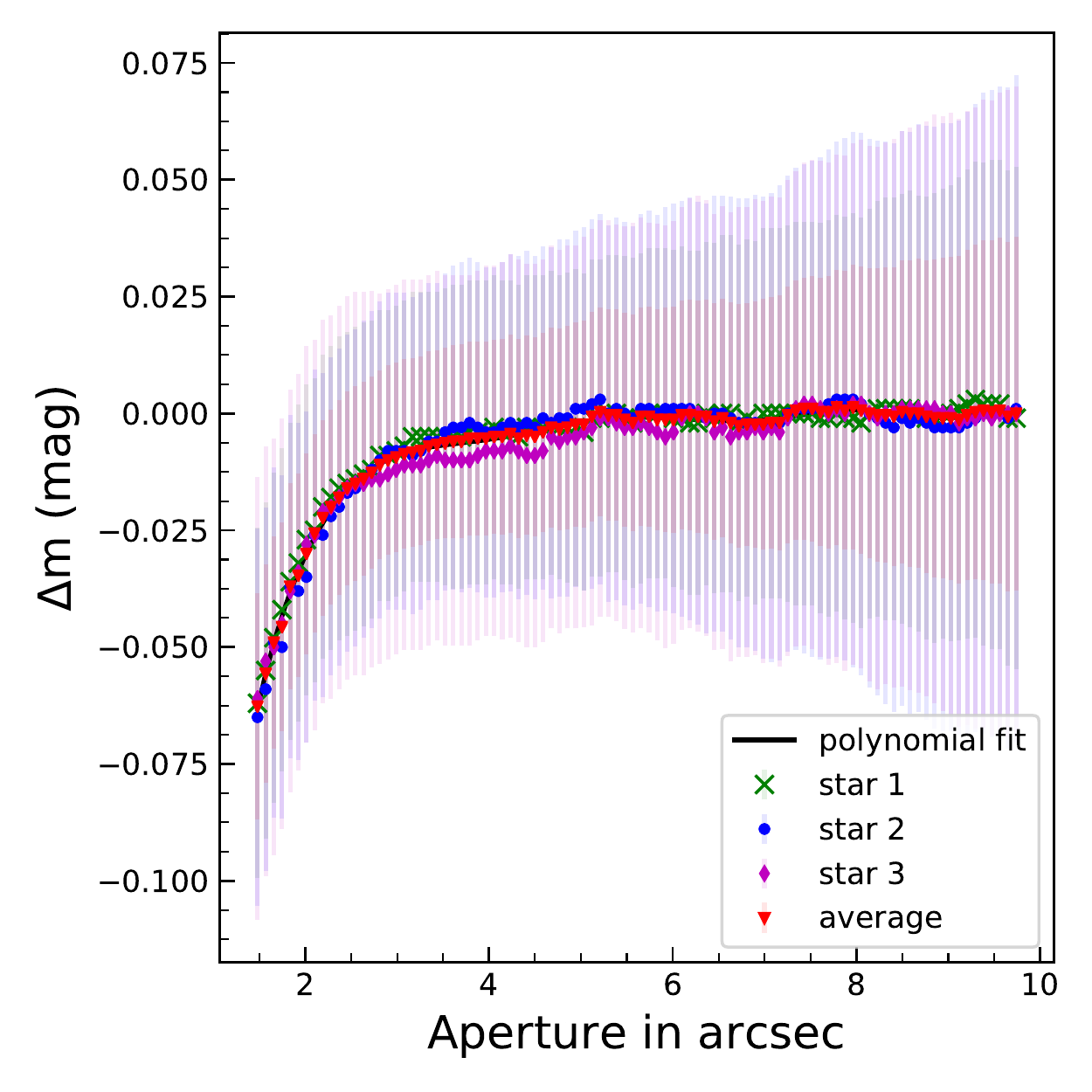}}
\caption{S/N vs. aperture (top). Differential magnitude between successive apertures vs. aperture size (bottom). The green, blue, purple and the red points represent differential magnitudes for comparison star 1, star 2, star 3 and average of those, respectively. The black solid line is the best polynomial fit curve for the average differential magnitudes.}
\label{fig:fig-s_n}
\end{figure}

\begin{table*}
\caption{Results of photometry. The fluxes and their uncertainties in 
different photometric bands are in units of $\mathrm{10^{-26} \, erg \, s^{-1} \, cm^{-2} \, Hz^{-1}}$. JD in days is given as JD $-$ 2457900.00.}
\label{tab:table-1}
\small
\setlength{\tabcolsep}{4pt}
\begin{tabular}{llllllllllllllll} \hline
JD(B)         & F(B)  & $\sigma_B$ & JD(V) & F(V) & $\sigma_V$ & $\alpha$ & JD(J) & F(J) & $\sigma_J$ & JD(H) & F(H) & $\sigma_H$ & JD(K) & F(K) & $\sigma_K$ \\
\hline

$-$ & $-$ & $-$ & 61.2132 & 0.475 & 0.012 & $-$ & 61.2006 & 3.437 & 0.171 & 61.1952 & 4.284 & 0.172 & 61.1898 & 4.769 & 0.263 \\
$-$ & $-$ & $-$ & 83.1235 & 0.577 & 0.012 & $-$ & 83.0986 & 4.514 & 0.164 & 83.0931 & 4.589 & 0.164 & 83.0869 & 5.323 & 0.214 \\
$-$ & $-$ & $-$ & 89.1020 & 0.499 & 0.011 & $-$ & 89.0899 & 3.436 & 0.141 & 89.0846 & 4.481 & 0.166 & 89.0790 & 4.853 & 0.208 \\
90.2687 & 0.589 & 0.009 & 90.2637 & 0.526 & 0.012 & 0.508 & 90.1166 & 3.548 & 0.136 & 90.1111 & 4.626 & 0.162 & 90.1056 & 4.953 & 0.203 \\
109.0943 & 0.775 & 0.009 & 109.0821 & 0.699 & 0.011 & 0.467 & 109.1157 & 3.635 & 0.152 & 109.1104 & 4.803 & 0.174 & 109.1024 & 5.445 & 0.237 \\
113.0778 & 0.629 & 0.008 & 113.0743 & 0.580 & 0.011 & 0.361 & 118.0661 & 3.752 & 0.175 & 118.0606 & 4.858 & 0.177 & 118.0547 & 5.466 & 0.233 \\
118.0949 & 0.674 & 0.009 & 118.0811 & 0.606 & 0.011 & 0.478 & 121.2985 & 4.258 & 0.175 & 121.3043 & 5.214 & 0.186 & 121.3095 & 6.104 & 0.248 \\
121.2953 & 0.681 & 0.009 & 121.2823 & 0.593 & 0.010 & 0.623 & 123.0859 & 3.678 & 0.162 & 123.0921 & 5.063 & 0.181 & 123.0983 & 5.631 & 0.235 \\
123.0754 & 0.676 & 0.009 & 123.0688 & 0.644 & 0.011 & 0.219 & 128.0531 & 3.952 & 0.170 & 128.0531 & 4.936 & 0.179 & 128.0641 & 6.173 & 0.252 \\
128.0900 & 0.716 & 0.009 & 128.0763 & 0.633 & 0.011 & 0.556 & 135.0515 & 3.920 & 0.162 & 135.0656 & 5.156 & 0.184 & 135.0710 & 6.050 & 0.249 \\
135.0910 & 0.641 & 0.008 & 135.0858 & 0.619 & 0.011 & 0.155 & 138.1881 & 4.195 & 0.183 & 138.1940 & 5.010 & 0.178 & 138.1996 & 5.955 & 0.243 \\
138.1768 & 0.597 & 0.008 & 138.1716 & 0.557 & 0.011 & 0.318 & 140.0493 & 3.934 & 0.171 & 140.0571 & 4.970 & 0.296 & 143.1466 & 5.934 & 0.245 \\
143.1754 & 0.607 & 0.008 & 143.1723 & 0.546 & 0.011 & 0.479 & 143.1576 & 4.128 & 0.168 & 143.1576 & 4.885 & 0.174 & 145.1724 & 5.970 & 0.254 \\
145.1519 & 0.590 & 0.008 & 145.1388 & 0.557 & 0.010 & 0.259 & 145.1555 & 4.281 & 0.182 & 145.1666 & 4.745 & 0.177 & 146.1533 & 6.050 & 0.253 \\
146.1875 & 0.592 & 0.008 & 146.1770 & 0.567 & 0.010 & 0.194 & 146.1643 & 4.077 & 0.176 & 146.1589 & 4.979 & 0.176 & 148.1671 & 5.909 & 0.240 \\
148.1609 & 0.618 & 0.008 & 148.1557 & 0.599 & 0.010 & 0.143 & 148.1777 & 4.062 & 0.164 & 148.1726 & 4.941 & 0.174 & 151.2010 & 5.958 & 0.343 \\
151.1697 & 0.584 & 0.008 & 151.1627 & 0.577 & 0.010 & 0.059 & 151.1863 & 4.016 & 0.381 & 151.1946 & 5.389 & 0.263 & 156.0789 & 7.180 & 0.282 \\
156.1047 & 0.578 & 0.009 & 156.1011 & 0.511 & 0.010 & 0.554 & 156.0598 & 4.763 & 0.205 & 156.0732 & 6.332 & 0.220 & 168.0090 & 5.897 & 0.234 \\
163.0652 & 0.600 & 0.009 & 163.0527 & 0.554 & 0.010 & 0.352 & 168.0206 & 3.859 & 0.167 & 168.0149 & 4.665 & 0.168 & 169.1878 & 5.847 & 0.253 \\
168.0448 & 0.654 & 0.008 & 168.0346 & 0.595 & 0.011 & 0.426 & 169.1765 & 3.679 & 0.166 & 169.1822 & 4.835 & 0.190 & 172.0439 & 6.183 & 0.256 \\
169.1735 & 0.612 & 0.008 & 169.1642 & 0.583 & 0.010 & 0.218 & 172.0249 & 4.107 & 0.193 & 172.0363 & 5.099 & 0.187 & 178.0571 & 5.417 & 0.231 \\
172.0607 & 0.608 & 0.009 & 172.0555 & 0.532 & 0.011 & 0.605 & 178.0436 & 3.454 & 0.155 & 178.0510 & 4.695 & 0.176 & 183.1151 & 6.044 & 0.258 \\
178.0323 & 0.350 & 0.008 & 178.0257 & 0.387 & 0.010 & -0.449 & 183.1041 & 3.685 & 0.160 & 183.1096 & 4.945 & 0.184 & 188.1144 & 5.172 & 0.245 \\
183.0956 & 0.452 & 0.008 & 183.0766 & 0.467 & 0.010 & -0.145 & 188.1011 & 3.690 & 0.168 & 188.1074 & 4.333 & 0.172 & 225.0356 & 5.360 & 0.236 \\
188.1374 & 0.221 & 0.008 & 188.1263 & 0.272 & 0.010 & -0.930 & 225.0227 & 3.976 & 0.184 & 225.0297 & 4.753 & 0.185 & 264.4761 & 5.026 & 0.228 \\
225.0555 & 0.325 & 0.008 & 225.0455 & 0.328 & 0.010 & -0.046 & 264.4650 & 3.678 & 0.176 & 264.4708 & 4.694 & 0.185 & 269.4794 & 5.067 & 0.244 \\
264.4490 & 0.489 & 0.008 & 264.4438 & 0.471 & 0.011 & 0.171 & 269.4901 & 3.613 & 0.259 & 269.4849 & 4.461 & 0.201 & 287.4277 & 5.448 & 0.235 \\
269.4742 & 0.506 & 0.008 & 269.4643 & 0.495 & 0.011 & 0.100 & 287.4387 & 3.655 & 0.165 & 287.4332 & 4.633 & 0.168 & 293.3966 & 5.860 & 0.253 \\
287.4235 & 0.577 & 0.008 & 287.4170 & 0.509 & 0.011 & 0.557 & 293.4109 & 3.807 & 0.174 & 293.4053 & 4.792 & 0.177 & 300.3659 & 5.370 & 0.232 \\
293.3921 & 0.640 & 0.008 & 293.3789 & 0.580 & 0.011 & 0.440 & 300.3530 & 3.543 & 0.182 & 300.3603 & 4.750 & 0.182 & 312.3806 & 5.073 & 0.215 \\
300.4129 & 0.540 & 0.008 & 300.4077 & 0.481 & 0.010 & 0.514 & 312.3697 & 3.557 & 0.172 & 312.3751 & 4.435 & 0.165 & 324.3247 & 4.968 & 0.233 \\
312.3519 & 0.482 & 0.009 & 312.3450 & 0.426 & 0.011 & 0.556 & 319.3554 & 3.363 & 0.164 & 319.3516 & 4.113 & 0.191 & 330.4025 & 4.861 & 0.212 \\
319.3354 & 0.357 & 0.008 & 319.3393 & 0.389 & 0.010 & -0.384 & 324.3136 & 3.403 & 0.170 & 324.3193 & 4.563 & 0.176 & 336.3635 & 5.038 & 0.223 \\
330.3760 & 0.421 & 0.008 & 330.3728 & 0.398 & 0.010 & 0.254 & 330.3907 & 3.391 & 0.154 & 330.3968 & 4.115 & 0.158 & 344.3510 & 3.362 & 0.241 \\
336.3370 & 0.366 & 0.008 & 336.3294 & 0.343 & 0.010 & 0.298 & 336.3527 & 3.118 & 0.153 & 336.3580 & 3.769 & 0.148 & 349.4212 & 4.696 & 0.207 \\
344.4617 & 0.334 & 0.009 & 344.4566 & 0.416 & 0.011 & -0.986 & 344.4466 & 2.650 & 0.181 & 344.4413 & 3.014 & 0.179 & 355.4278 & 4.926 & 0.228 \\
349.4072 & 0.530 & 0.008 & 349.4019 & 0.478 & 0.010 & 0.467 & 349.4331 & 3.190 & 0.160 & 349.4275 & 4.105 & 0.162 & 361.3132 & 4.847 & 0.219 \\
355.3493 & 0.622 & 0.008 & 355.3386 & 0.573 & 0.011 & 0.369 & 355.4170 & 3.019 & 0.159 & 355.4223 & 4.091 & 0.162 & 377.2408 & 5.145 & 0.247 \\
361.2995 & 0.615 & 0.008 & 361.3054 & 0.550 & 0.011 & 0.500 & 361.3274 & 3.444 & 0.169 & 361.3216 & 4.319 & 0.166 & 384.2644 & 5.676 & 0.268 \\
372.3871 & 0.752 & 0.009 & 372.3810 & 0.598 & 0.011 & 1.029 & 372.3407 & 3.640 & 0.164 & 372.3256 & 4.780 & 0.178 & 389.1509 & 5.492 & 0.269 \\
377.2167 & 0.731 & 0.009 & 377.2085 & 0.659 & 0.011 & 0.463 & 377.2276 & 3.589 & 0.177 & 377.2351 & 4.322 & 0.176 & 394.2799 & 5.423 & 0.252 \\
384.3065 & 0.782 & 0.009 & 384.3013 & 0.702 & 0.012 & 0.485 & 384.2538 & 3.640 & 0.187 & 384.2588 & 4.672 & 0.192 & 404.1516 & 5.110 & 0.268 \\
389.1414 & 0.753 & 0.010 & 389.1275 & 0.712 & 0.012 & 0.251 & 389.1633 & 3.851 & 0.196 & 389.1578 & 4.972 & 0.200 & 445.2798 & 4.626 & 0.312 \\
394.2751 & 0.751 & 0.010 & 394.2650 & 0.647 & 0.012 & 0.670 & 394.2913 & 3.567 & 0.177 & 394.2856 & 4.681 & 0.184 & 482.0729 & 4.386 & 0.204 \\
404.1918 & 0.422 & 0.008 & 404.1883 & 0.440 & 0.011 & -0.190 & 404.1678 & 2.890 & 0.253 & 404.1572 & 4.500 & 0.196 & 504.1383 & 6.320 & 0.269 \\
445.2936 & 0.131 & 0.007 & 445.2905 & 0.163 & 0.010 & -0.967 & 445.2718 & 2.763 & 0.156 & 445.2757 & 3.864 & 0.167 & $-$ & $-$ & $-$ \\ 
482.1024 & 0.701 & 0.009 & 482.0976 & 0.655 & 0.011 & 0.311 & 482.0836 & 2.986 & 0.137 & 482.0782 & 3.858 & 0.147 & $-$ & $-$ & $-$ \\
504.1248 & 0.502 & 0.008 & 504.1176 & 0.488 & 0.011 & 0.122 & 504.1508 & 3.436 & 0.150 & 504.1449 & 4.819 & 0.173 & $-$ & $-$ & $-$ \\

\hline
\end{tabular}
\end{table*}
\subsection{Correction for the host-galaxy contribution}
The COG method used to obtain the magnitude of the comparison stars
could not be used for the target AGN as both the B and V band images
have a prominent host galaxy contribution (see Fig. \ref{fig:fig-3} left). 
The contribution of the host galaxy in both the B and V band needs to be 
subtracted to get the true flux from the AGN in B and V bands.  For that, we 
used the two dimensional image-decomposition code 
GALFIT \citep{2002AJ....124..266P}. For one good epoch of observation
in B and V bands, we fitted a S\'ersic profile
and an edge-on disk profile component to the data. For generating the model PSF we used a point source present
close to the target source. An example of the observed AGN, the  modeled galaxy and 
the residual image that contains the AGN (after subtraction of the 
model galaxy) for V band are shown in the middle and right panels of 
Fig. \ref{fig:fig-3}. Light curves of the AGN, ideally, can be generated from 
the photometry of the residual AGN image obtained after GALFIT. However,
it is difficult in practice due to the poor signal-to-noise (S/N) ratio in many 
epochs of data. Therefore, to get the light curve of the AGN devoid of the host 
galaxy, we followed the following approach. For each epoch of observation, we 
did aperture photometry of the target Z229$-$15 at an aperture radius equal to the aperture used for the comparison stars of that epoch. The derived flux ($F_{\mathrm{total}}$) contains the
light from the AGN ($F_{\mathrm{AGN}}$) and the host galaxy ($F_{\mathrm{host}}$). Now to 
remove the host galaxy light we did aperture photometry on the modeled galaxy 
image (obtained from GALFIT and without the AGN)  at different concentric 
aperture sizes. The flux of the galaxy as a function of radii from its center 
was generated and then modeled as a polynomial. This is shown in 
Fig.  \ref{fig:fig-4}. Once the functional form of the galaxy light 
distribution was obtained, we found the contribution of
galaxy at the radius used for the photometry of Z229$-$15 for that epoch. This was then subtracted from the total flux to get the light curve of 
the AGN as
\begin{equation}
F_{\mathrm{AGN}} = F_{\mathrm{total}} - F_{\mathrm{host}}
\end{equation}
The above was carried out for each epoch of observation to arrive at the final
B and V band light curves of the AGN.  Finally, the observed instrumental 
magnitudes were converted into apparent magnitudes using differential 
photometry relative to the two comparison stars having brightness 
similar to the AGN in the field, and, whose apparent magnitudes were taken 
from \citet{2011ApJ...732..121B} and \citet{2011MNRAS.416..403F} for V and B band, respectively. These comparison stars are found to be non-variable
during our monitoring period (see Fig. \ref{fig:fig-2}).   The obtained 
apparent magnitudes were corrected for Galactic extinction from 
the NASA/IPAC Extragalactic data base (NED)
\footnote{\url{https://nedwww.ipac.caltech.edu}}. These magnitudes were then 
converted into fluxes using the zero points taken from 
\citet{1979PASP...91..589B}.

\subsection{Infrared photometry}

The images acquired in the NIR J, H and $\mathrm{K_s}$ bands have very poor 
S/N compared to the optical B and V band images. Also,
the PSF is found to change across the image frames. Similar to the 
optical band we tried to obtain the magnitude of the comparison star and
Z229$-$15 (as the host galaxy is not seen) using
the COG method.  However, we found the COG not to smoothly merge with the 
background and instead showed many wiggles (see Fig. \ref{fig:fig-5}). 
Therefore, to obtain the total fluxes from the comparison stars and the 
AGN  we followed a two step 
approach. We carried out aperture photometry on a sequence of circular
apertures. We calculated S/N at each of the aperture and plotted S/N as
a function of aperture to find the aperture at which the S/N is 
maximum (see Fig. \ref{fig:fig-s_n}, top panel). We used that aperture to find the magnitude of the objects. 
The aperture at which the S/N is maximum changes from  epoch to epoch. Due to poor S/N, the optimum aperture obtained by this method is much smaller than the FWHM and therefore, the flux obtained on 
the comparison stars and the AGN at the optimum aperture is always an 
underestimation. 

Therefore, to get the true brightness of the comparison 
stars and the AGN we need to apply aperture correction to offset for the 
missing flux obtained at smaller apertures. We estimated the aperture 
correction using a mean differential curve of growth (DCOG) analysis. 
There are three point sources in the observed image frames in the NIR
bands.  For each of these stars we calculated the magnitude difference between consecutive apertures, i.e $\Delta m = m_{i + i} - m_i$, where $i$ refers to
the aperture radius, and plotted $\Delta m$ against aperture radii as shown 
in Fig. \ref{fig:fig-s_n} (bottom panel). A mean DCOG is then obtained by taking the 
average of the DCOG obtained for these three stars. A polynomial is 
fit to the mean DCOG. The difference between the $\Delta m$ at the FWHM and the $\Delta m$ at the optimum aperture obtained from the best-fit polynomial was applied
as aperture correction to each of the comparison stars and
the AGN to get their actual brightness.  We note that the pattern of the NIR 
light curves before and after the correction looks nearly identical, however, there
is a minor advantage (though not much significant) in using the correction 
factor obtained from 
the DCOG. For example, the standard deviation of the DLC between Star 1 and Star 2 before
applying the correction factor is 0.096 mag in the J band and it reduces to 0.087 mag
after applying the correction factors. In the H and $\mathrm{K_s}$ bands, the standard deviation of 
the DLC between Star 1 and Star 2 are similar before and after application of the correction
factors.
The final NIR magnitudes  were obtained by carrying out differential 
photometry of the AGN relative to the three comparison stars, the 
standard magnitudes of which were taken from SIMBAD \footnote{\url{http://simbad.u-strasbg.fr/simbad/}} data base. They were corrected
for Galactic extinction using the values taken from NED. Final NIR 
light curves in flux units were obtained using the zero points from
\citet{1998A&A...333..231B}.  

The NIR light curves obtained by the procedure outlined above will have 
contribution from the torus, the host galaxy, and the infrared emission coming from the accretion disk (AD). The observed infrared radiation can be written as
\begin{equation}
F_{\mathrm{obs}} = F_{\mathrm{dust}} + F_{\mathrm{host}} + F_{\mathrm{AD}}
\end{equation}
Therefore, to generate the infrared light curves that contain only the 
reprocessed optical continuum from the accretion disk by the dust torus 
($F_{\mathrm{dust}}$), we need to subtract the contribution to the NIR from the 
host galaxy ($F_{\mathrm{host}}$) and the AD ($F_{\mathrm{AD}}$).  However, as the infrared images have poor S/N and as 
the AGN is only visible in the NIR images, we did not attempt to correct for 
$F_{\mathrm{host}}$ and instead aimed to correct only for the contribution of $F_{\mathrm{AD}}$ 
to the observed NIR emission.

\subsection{Subtraction of the AD component of the NIR flux}

The observed NIR emission has contribution from the AD.
\citep{2006ApJ...652L..13T, 2008Natur.454..492K, 2011MNRAS.415.1290L}. This AD contamination to the NIR fluxes makes the derived lag between 
the optical and NIR continuum shorter than the actual 
lag \citep{2014ApJ...788..159K}. So to get the actual time lag, between the optical and NIR flux variations, the contribution of the AD 
to the observed NIR fluxes needs to be removed. We estimated the 
AD contribution to the NIR fluxes by considering a power-law 
spectrum of the AD \citep{2014ApJ...788..159K} and written
as 

\begin{equation}
f_{\mathrm{NIR}}^{\mathrm{AD}}(t) = f_V(t) \left(\frac{\nu_{\mathrm{NIR}}}{\nu_{V}}\right)^{\alpha}
\label{eq:eq_ad}
\end{equation}   

where $f_{\mathrm{NIR}}^{\mathrm{AD}}(t)$ and $f_V(t)$ represent the accretion disk component of the NIR flux and the V band flux at time t, $\nu_V$ and 
$\nu_{\mathrm{NIR}}$ are the effective frequencies of V and the  NIR (J, H, $\mathrm{K_{s}}$) bands, respectively, and $\alpha$ is the power-law index.

\begin{figure}
\resizebox{8cm}{8cm}{\includegraphics{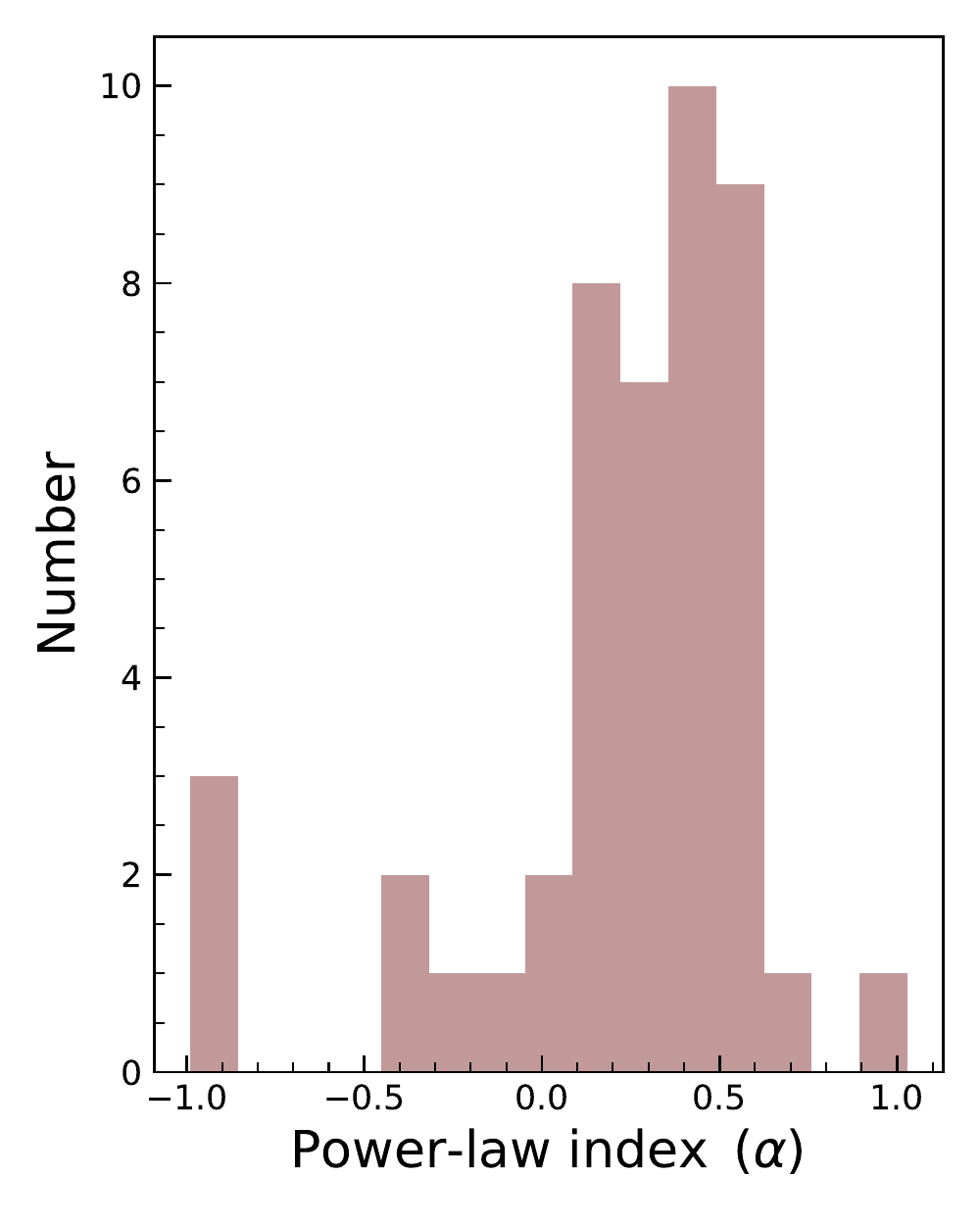}} 
\caption{Distribution of the spectral index $\alpha$.}
\label{fig:fig-6}
\end{figure}

\begin{figure}
\resizebox{8cm}{10cm}{\includegraphics{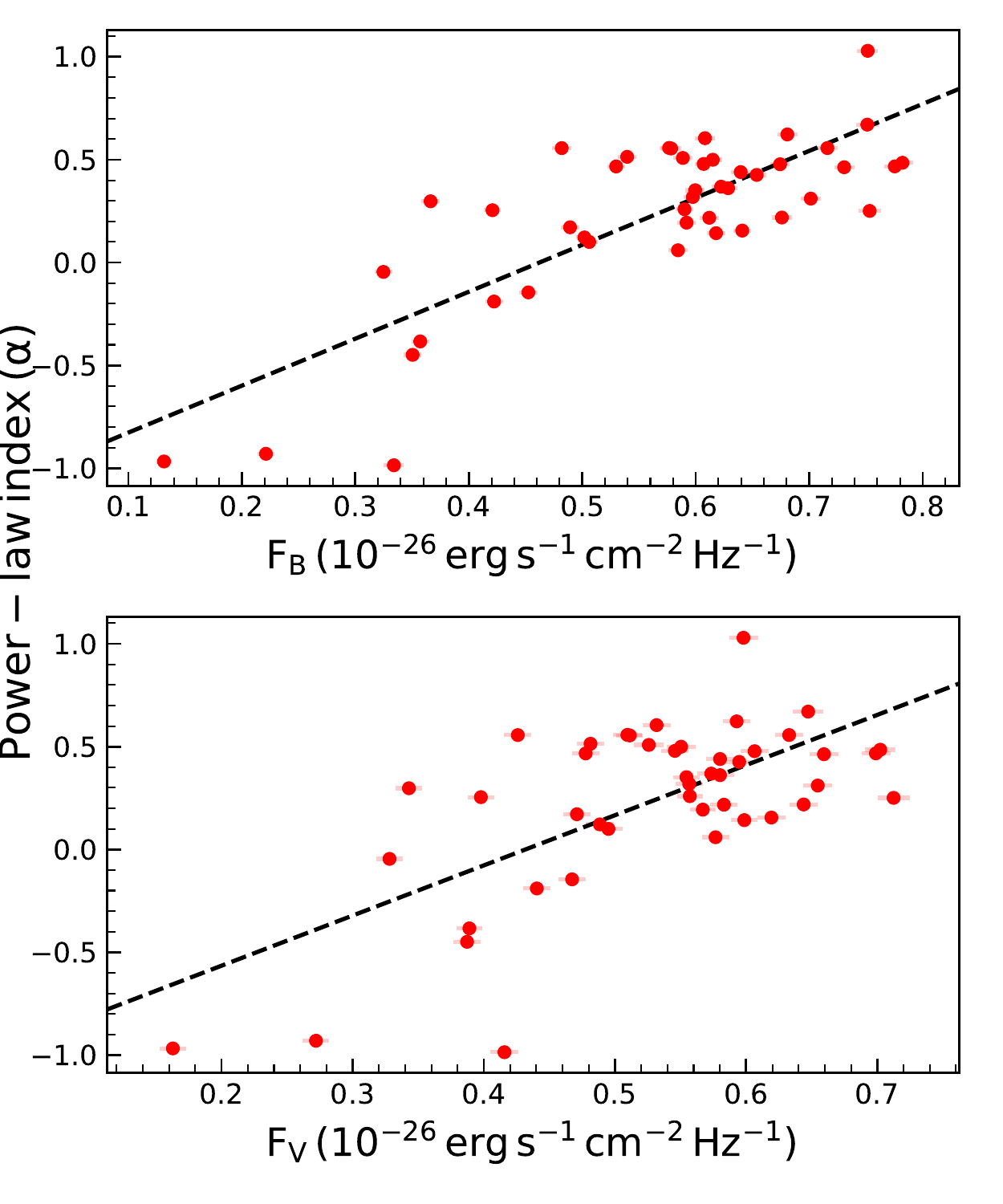}} 
\caption{ Dependency of $\alpha$ on B (top) and V band (bottom) flux.}
\label{fig:fig-alpha}
\end{figure}

 AGN do show spectral variations with their brightness 
\citep{2011A&A...525A..37M}. They are known to show a bluer when brighter behaviour \citep{2014ApJ...783...46K} which suggests that AGN have a time dependent power-law index $\alpha$. Therefore, each epoch of NIR observation
has to be corrected for the AD emission component using the $\alpha$
evaluated for each epoch. At any given epoch, the observations in the optical and each of the NIR bands were typically obtained within 300 sec of one another, and therefore, they were considered as nearly simultaneous to remove
the contribution of AD to each of the NIR bands.
We estimated $\alpha$ for each epoch of observation using the 
near-simultaneous observations in the optical 
B and V bands making use of the relation given below 
\citep{2019BSRSL..88..158M}  

\begin{equation}
\alpha = \frac{ln\big(f_B/f_V\big)}{ln\big(\nu_B/\nu_V\big)}
\end{equation}

where, $f_B$ and $f_V$ are the flux densities in B and V band, 
respectively, while $\nu_B$ and $\nu_V$ are the frequencies in B and 
V band. The distribution of $\alpha$ obtained between B and V bands for all the epochs of observations is shown in Fig. \ref{fig:fig-6}. The
estimated values of $\alpha$ range between $-$0.99 and 1.03. We found a median $\alpha$ of 0.318 with a standard deviation of 0.418. This is similar to the 
value of $\alpha$ = 1/3 expected in the standard accretion model
\citep{1973A&A....24..337S}. 

 The variation of $\alpha$ with both the B and V band brightness of the source
is shown in Fig. \ref{fig:fig-alpha}. We found $\alpha$ to be correlated significantly 
with the B and V band brightness with a bluer when brighter trend.
Linear least squares fit to the data gave Spearman rank correlation coefficients 
of 0.598 and 0.442 with the  probability (p) of no correlation of $10^{-5}$ and 0.002 for
B and V bands, respectively.  The $\alpha$ value obtained for each
epoch was used  to calculate the epoch-wise values of  $F_{\mathrm{AD}}$ using 
Equation \ref{eq:eq_ad}.  The calculated values of $F_{\mathrm{AD}}$  were then subtracted from the observed $F_{\mathrm{obs}}$ values to get the dust contribution. 
The error in the flux values in the different bands were obtained via error propagation. The results of the photometry are given in  Table \ref{tab:table-1}.


\section{Analysis}
\subsection{Flux variability}
\begin{figure}
\includegraphics[scale=0.8]{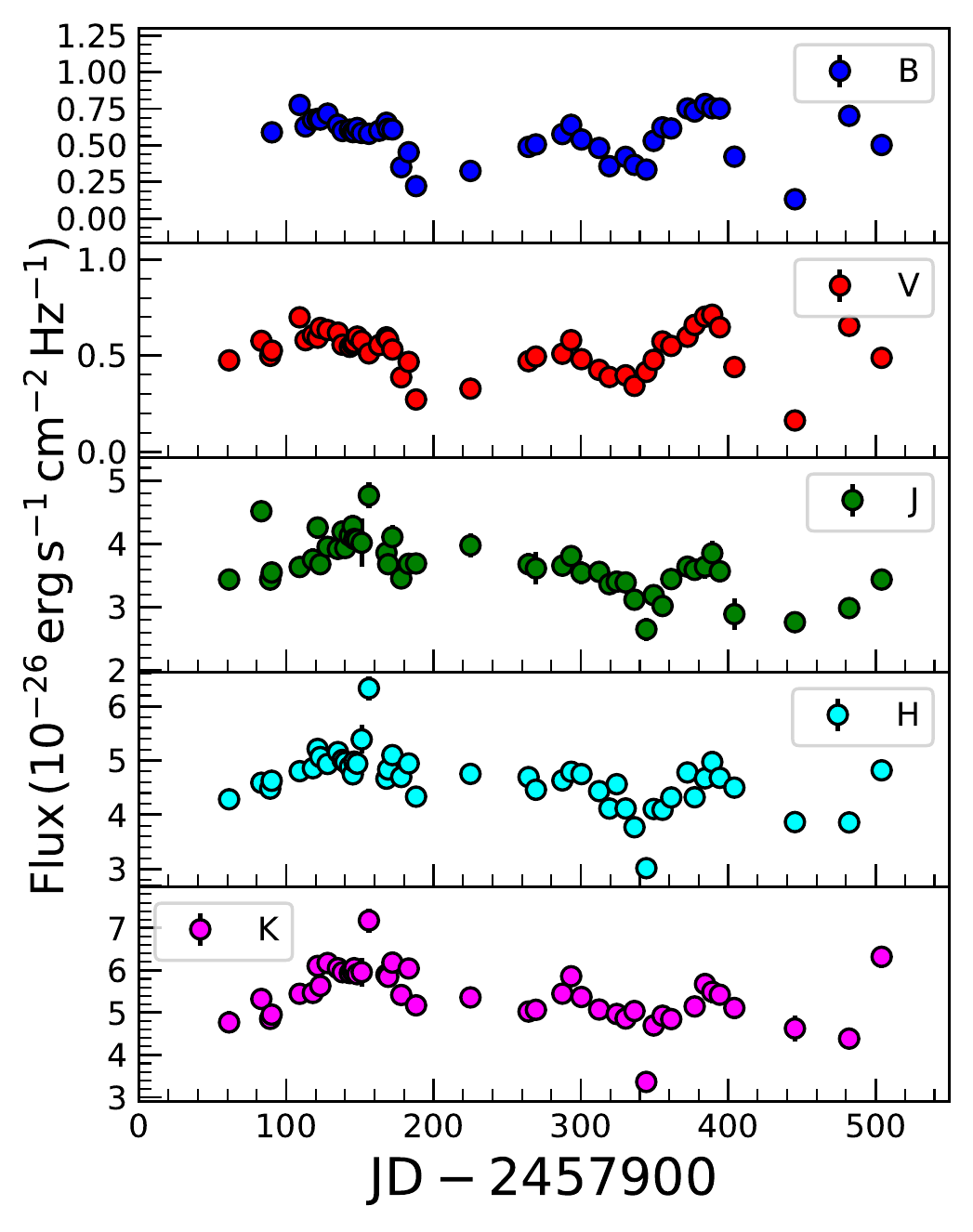}
\caption{The light curves of Z229$-$15 in optical B, V and NIR J, H, $\mathrm{K_s}$ bands for the period from 2017 July to 2018 December. The accretion disk component was subtracted from all the NIR light curves using $\alpha$ measured at individual epoch.}
\label{fig:fig-9}
\end{figure}

We show in Fig. \ref{fig:fig-9} the final light curves of Z229$-$15 in  
 the optical B and V bands and  the NIR J, H and $\mathrm{K_s}$ bands. We characterised the flux variability of the source using normalized excess 
variance (F$_{\mathrm{var}}$; \citealt{2002ApJ...568..610E,2003MNRAS.345.1271V,2017MNRAS.466.3309R}).   
The uncertainties in the $F_{\mathrm{var}}$ values were calculated following \citet{2002ApJ...568..610E}. The results of the flux variability
analysis of the source are given in  Table \ref{tab:table-2}. Here, $<f>$ and $\sigma$
are the mean and standard deviation of the light curves and $R_{\mathrm{max}}$ is 
the ratio between the maximum and minimum flux in the light curve. We found
wavelength dependent variability with the amplitude of variability in the 
shorter wavelength (B band) larger than that at the longer wavelength (K band). 

\begin{figure}
\includegraphics[scale=0.62]{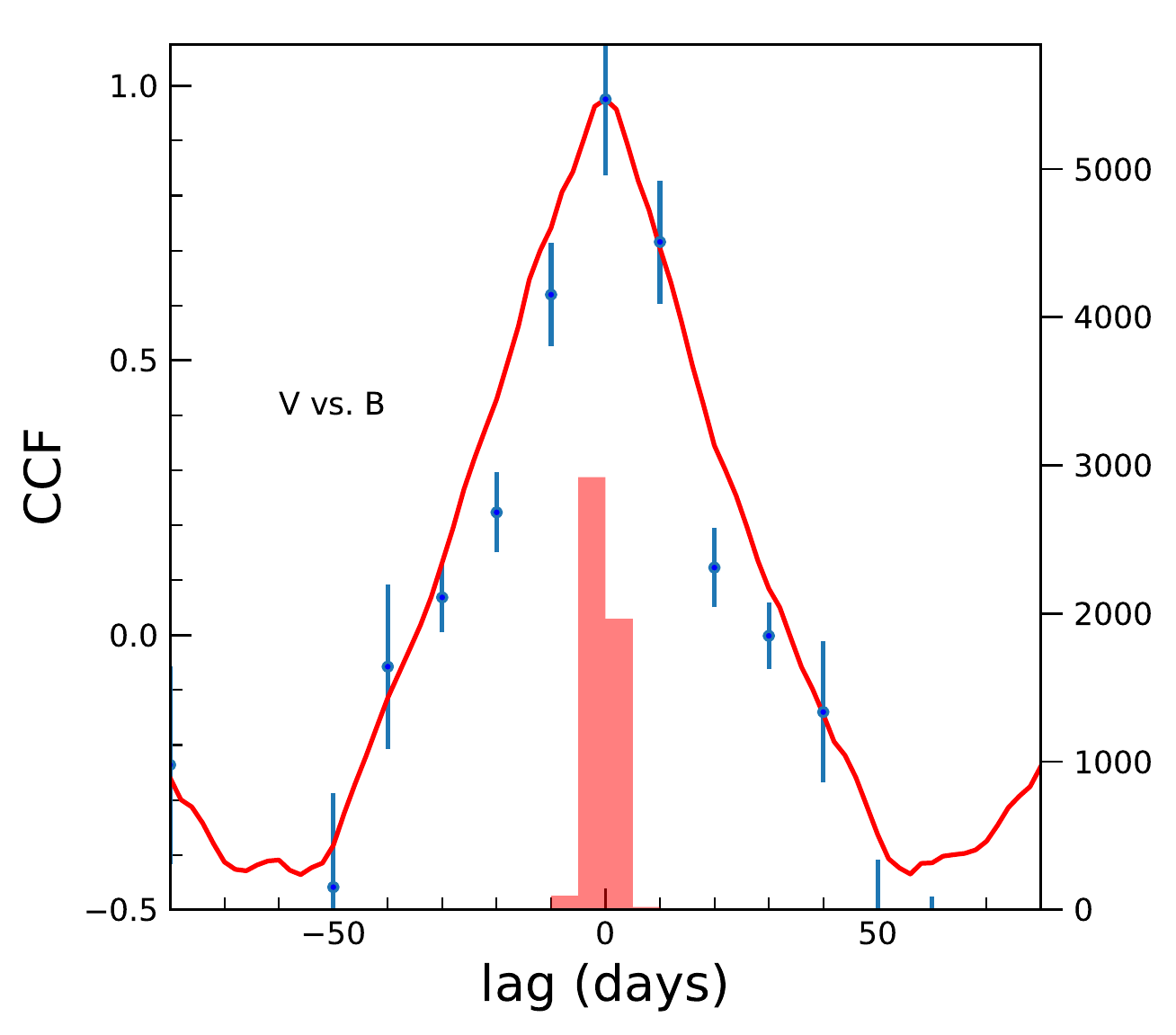}
\caption{Correlation between V and B band. The solid red line represents the 
	ICCF and points with error bars show the DCF. The distribution of  
	$\tau_{\mathrm{cent}}$  obtained from ICCF is also shown.}
\label{fig:fig-10}
\end{figure}

\begin{table}
\caption{Results of the analysis of variability in the observer's frame.
The mean values of fluxes ($<f>$) in various bands and their
associated standard deviations $\sigma$ are in units of
$\mathrm{10^{-26} \, erg \, s^{-1} \, cm^{-2} \, Hz^{-1}}$. $\lambda_{eff}$ is the effective wavelength in Angstroms.}
\label{tab:table-2}
\begin{tabular}{llcccl} \hline
Filter & $\lambda_{\mathrm{eff}}$ & $<f>$ & $\sigma$ & $F_{\mathrm{var}}$ & $R_{\mathrm{max}}$  \\
\\ \hline
B & 4363 & 0.56  & 0.15 & 0.263 $\pm$ 0.000 & 5.953  \\
V & 5448 &  0.53 & 0.11 & 0.213 $\pm$ 0.000 & 4.374  \\
J & 12200 &  3.66 & 0.43 & 0.106 $\pm$ 0.002 & 1.797  \\
H & 16300 &  4.64 & 0.50 & 0.100 $\pm$ 0.002 & 2.101  \\
$\mathrm{K_{s}}$ & 21900 & 5.46 & 0.61 & 0.102 $\pm$ 0.002 & 1.905  \\
\hline
\end{tabular}
\end{table}

\subsection{Lag between optical and NIR variations}

The light curves in the optical B, V and  the NIR J, H, 
$\mathrm{K_{s}}$ bands (see Fig. 
\ref{fig:fig-9}) and the results given in Table \ref{tab:table-1} indicate that 
Z229$-$15 is variable in the optical and NIR bands and therefore it is 
possible to estimate time lag if any between the flux variation
in different optical and NIR bands.

\begin{figure}
\includegraphics[scale=0.62]{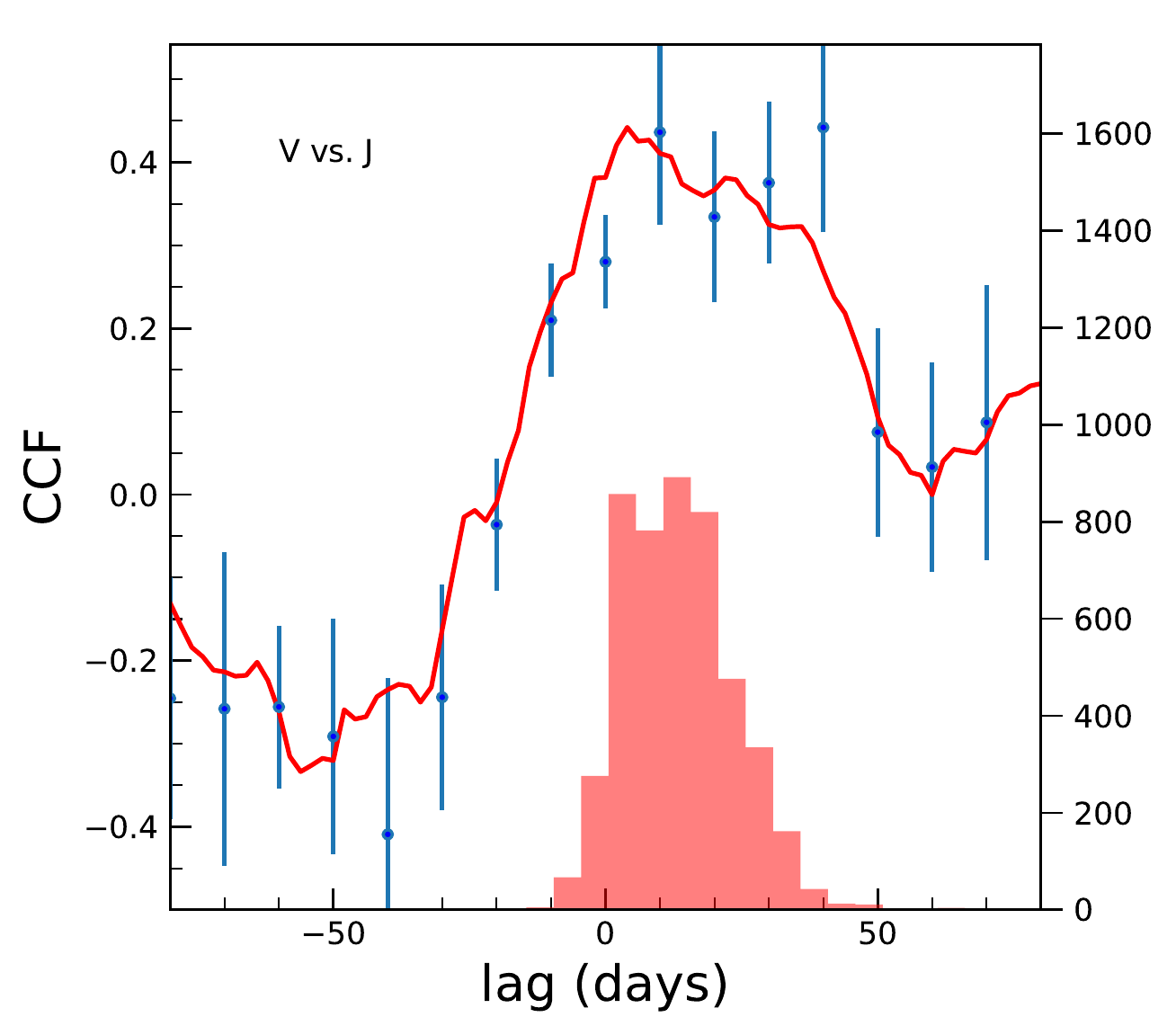}
\includegraphics[scale=0.62]{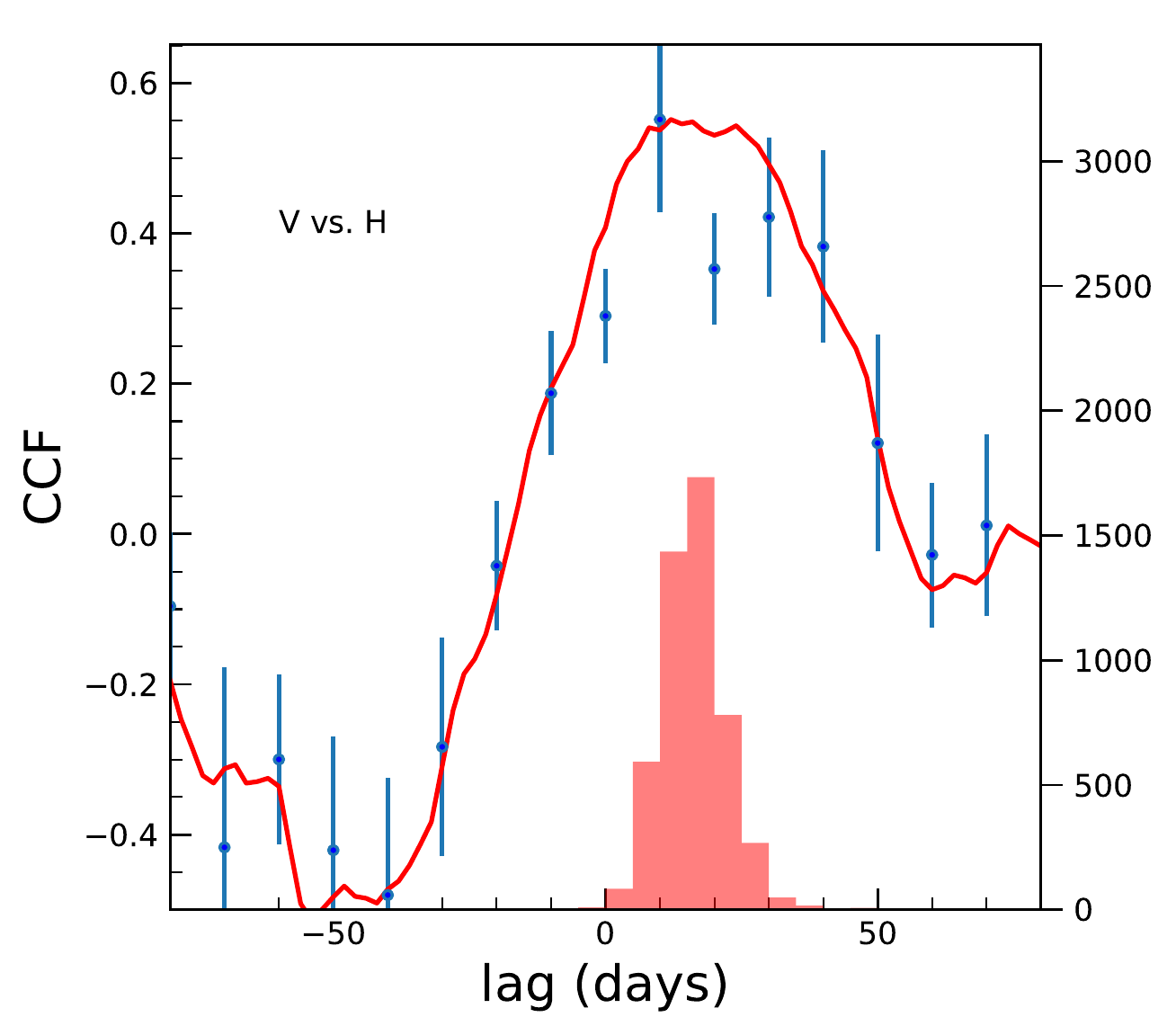}
\includegraphics[scale=0.62]{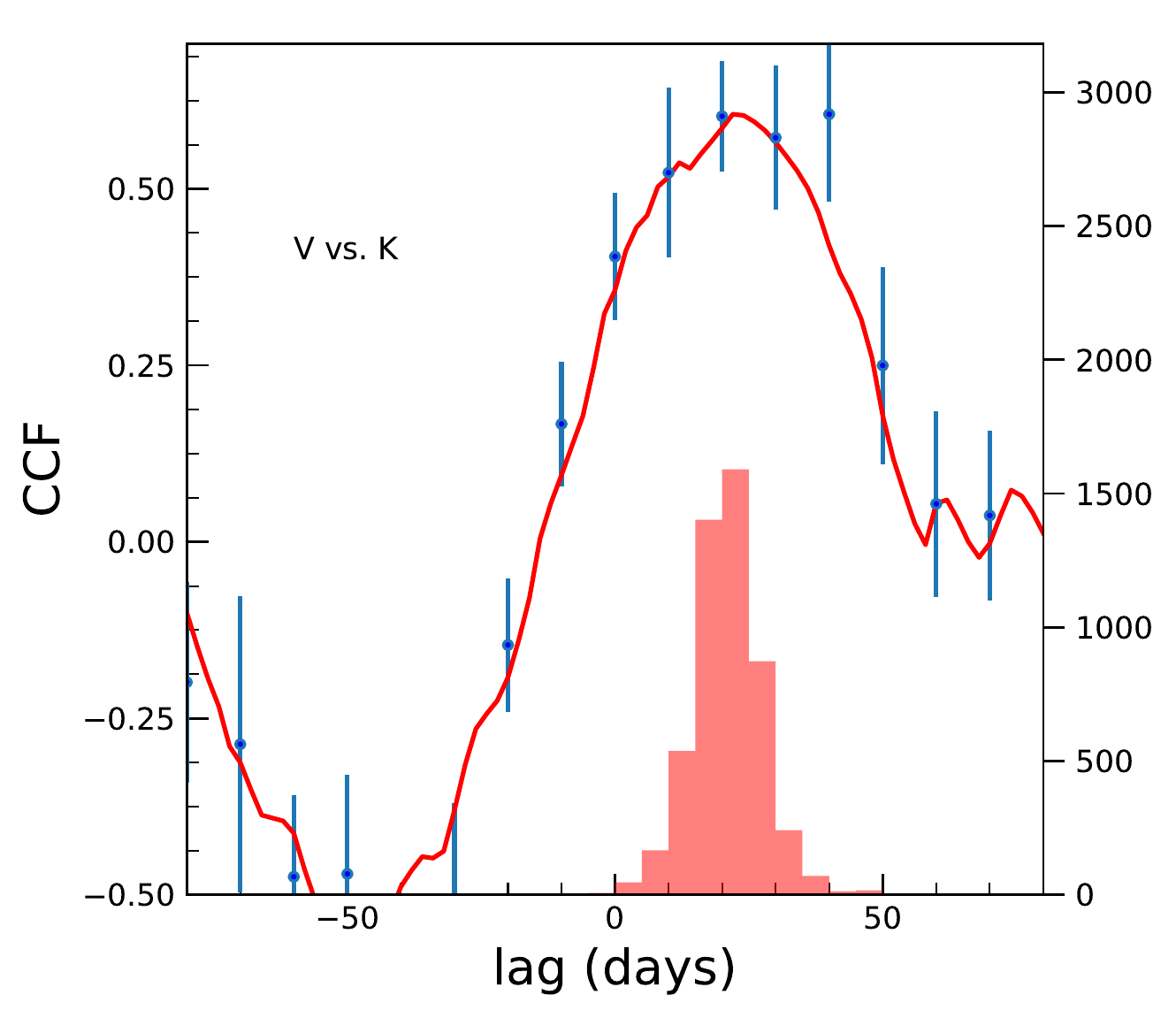}
\caption{The CCFs of V versus J (top panel), V versus H (middle panel),
and V versus $\mathrm{K_{s}}$ (bottom panel) are shown. The solid red line 
	represents the ICCF and the points with error bars show the DCF. 
	The distribution of $\tau_{\mathrm{cent}}$  obtained from ICCF 
	method} is also shown in each panel.
\label{fig:fig-11}
\end{figure}

\subsubsection{Cross-correlation analysis}

To estimate the time lag
between  the optical and the NIR flux variations, we used two well known 
cross correlation function (CCF) methods, 
namely the interpolated cross-correlation function 
(ICCF; \citealt{1986ApJ...305..175G,1987ApJS...65....1G}) and the discrete correlation function (DCF; \citealt{1988ApJ...333..646E}). For CCF analysis we used a time bin ($\tau$) of 5 days, as this is the typical sampling of
our multi-band light curves.  We show in 
Fig. \ref{fig:fig-10} the DCF (blue filled circles) and ICCF (red solid line)
obtained between the optical B and V bands. The CCF peaks at zero lag.
Similarly, the CCF for three other filter combinations, namely V/J, V/H and V/$\mathrm{K_s}$ bands calculated using both the ICCF and DCF methods, are shown in  Fig. \ref{fig:fig-11}. The CCFs peak at a lag different from zero,  and also, 
the pattern of the CCFs obtained by both the DCF and ICCF methods are similar. 
To estimate the lag from the cross-correlation function
we calculated the centroid $\tau_{\mathrm{cent}}$ of the CCF as 
\citep{1998PASP..110..660P}

\begin{equation}
\tau_{\mathrm{cent}} = \frac{\Sigma_i \, \tau_i CCF_i} {\Sigma_i \, CCF_i}
\end{equation}

For calculating $\tau_{\mathrm{cent}}$, we used all the points that are within
80$\%$ of the maximum of the CCF. To estimate the uncertainties in the 
calculated $\tau_{\mathrm{cent}}$,  we used a model independent Monte Carlo 
simulation based on flux randomization (FR) and random subset 
selection (RSS; \citealt{1998PASP..110..660P, 1999ApJ...526..579W, 2004ApJ...613..682P}). This
was repeated for 5000 times and the centroid of the lag was determined each time.
The cross-correlation centroid distribution (CCCD) and cross-correlation peak distribution (CCPD) were then constructed. The CCCDs obtained using the ICCF method are shown in 
Fig. \ref{fig:fig-10} and 
Fig. \ref{fig:fig-11}. As the CCCD is not Gaussian, the lag is taken as the median of the 
distribution and the lower and upper error
in the lag are that values at the 15.9 and 84.1 percentile of the distribution. This corresponds to 1 sigma error in the case of a Gaussian distribution.
The results of the CCF analysis are given in Table \ref{tab:table-3}. 
The centroid lags calculated using both ICCF and DCF are found to be 
consistent with each other within error bars. 
However, in all further analysis we consider
the CCCD lags obtained from ICCF method as it has comparatively smaller uncertainties than the DCF lag (see Table \ref{tab:table-3}).

The rest frame  time lags corrected for the redshift can be obtained by dividing the observed time lag by a factor of $(1 + z)$ as prescribed by \citet{2014ApJ...788..159K}. Between
V and $\mathrm{K_s}$ we obtained a rest frame time lag of  $20.36^{+5.82}_{-5.68}$ days, while between V and H and V and J, we found rest frame time lags of 
$15.63^{+5.05}_{-5.11}$ days and $12.52^{+10.00}_{-9.55}$ days, respectively. 
There is an indication that the lag obtained between V and $\mathrm{K_s}$ band is 
larger than V and H, which is larger than V and J. However the error bars are too large
to unambiguously argue for the presence of wavelength dependent lag. 
From these observations, the lags obtained between different wavelengths are consistent with each other.
As the inner radius of the dust torus is defined by the lag between V 
and $\mathrm{K_{s}}$ band which corresponds to the dust sublimation radius, 
we used the rest frame lag between V and $\mathrm{K_{s}}$ band 
of $20.36^{+5.82}_{-5.68}$ days to infer the inner edge of the 
dust torus in Z229$-$15, which is at a distance of 0.017 pc from
the central continuum source.



\begin{table}
\caption{Median values of time delays with errors in days from the CCCDs obtained using DCF and ICCF,  CCPD from ICCF and  \textsc{javelin} in the observer's frame.}
\label{tab:table-3}
\small
\setlength{\tabcolsep}{1pt}
\begin{tabular}{l c c r r} \hline
	Band     & DCF  & ICCF & CCPD from ICCF &   \textsc{Javelin}   \\
\hline

	&  & varying $\mathrm{\alpha}$   &  &  
\\ \hline
	V $-$ B     & $-0.24^{+3.90}_{-4.39}$ & $-0.87^{+1.94}_{-2.00}$  & $0.00^{+0.00}_{-2.00}$ &   $-0.57^{+1.03}_{-1.29}$  \\
	V $-$ J     & $22.33^{+16.58}_{-15.16}$ & $12.86^{+10.27}_{-9.81}$  & $10.00^{+14.00}_{-8.00}$ &  $23.82^{+0.87}_{-0.45}$  \\
	V $-$ H     & $18.16^{+14.55}_{-9.91}$ & $ 16.06^{+5.19}_{-5.25}$  & $14.00^{+12.00}_{-6.00}$ &   $23.28^{+0.60}_{-0.40}$  \\
	V $-$ $ \mathrm{K_s}$   & $25.42^{+13.49}_{-14.88}$ & $20.92^{+5.98}_{-5.84}$  & $22.00^{+6.00}_{-10.00}$ & $23.04^{+0.74}_{-0.44}$  \\

\hline

	&  & constant $\mathrm{\alpha = 0.1}$    &  &  \\
    
\hline

	V $-$ J     & $34.01^{+11.99}_{-14.76}$ & $27.96^{+7.07}_{-8.73}$  &  $28.00^{+8.00}_{-8.00}$ &  $27.81^{+1.68}_{-1.47}$  \\
	V $-$ H     & $29.67^{+11.07}_{-12.19}$ & $26.14^{+5.85}_{-5.12}$  &  $26.00^{+6.00}_{-4.00}$ &   $25.07^{+1.96}_{-0.80}$   \\ 
	V $-$ $\mathrm{K_s}$     & $34.80^{+8.69}_{-12.93}$ & $30.86^{+6.22}_{-4.95}$  & $30.00^{+8.00}_{-4.00}$ &   $24.19^{+1.92}_{-0.72}$  \\

\hline
\end{tabular}
\end{table}

\begin{figure*}
\includegraphics[scale=1.0]{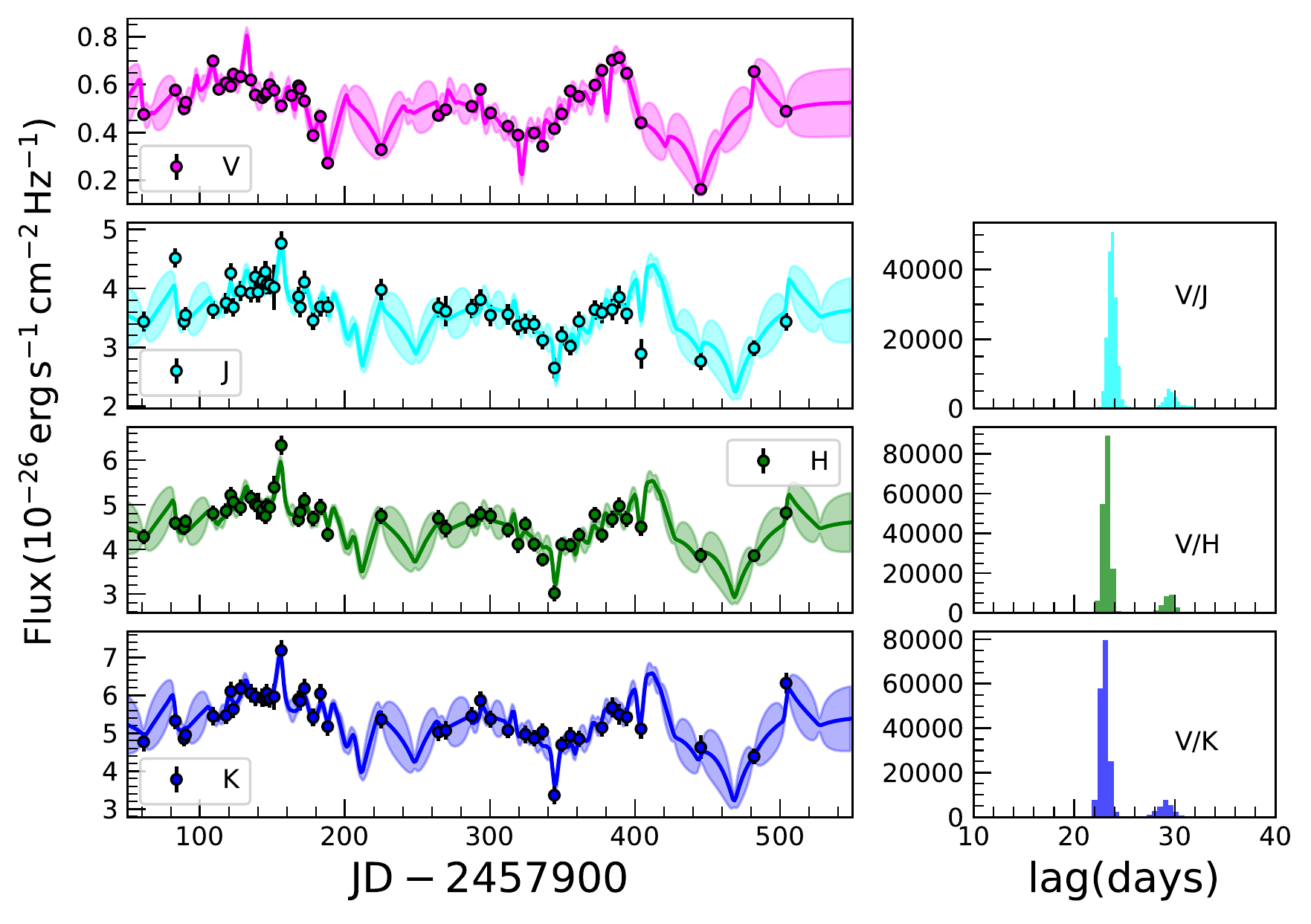}
\caption{Points with error-bars represent observed light curves, whereas the solid lines with shade are the recovered light curves using \textsc{javelin}. The accretion disk contamination in all the NIR light curves was subtracted using $\mathrm{\alpha}$ measured at individual epoch. The histograms represent the probability distribution of the time delay.  All the light curves are fitted simultaneously.}
\label{fig:fig-12}
\end{figure*}

\subsubsection{JAVELIN}
To calculate the time delays between the optical and different NIR (J, H, 
$\mathrm{K_s}$) bands, we also used  the \textsc{javelin} code, developed 
by \citet{2011ApJ...735...80Z, 2013ApJ...765..106Z}.  The driving continuum 
light curve was modeled using a damped random walk (DRW) process 
\citep[e.g.,][]{2009ApJ...698..895K} with two model parameters; amplitude and 
time scale of variability.  A top-hat response function was convolved with the 
driving continuum light curve to generate  the NIR continuum light curve. 
Therefore, the NIR light curves were the shifted, scaled, and smoothed version 
of the driving continuum light curve. The Markov chain Monte Carlo (MCMC) 
approach was used to find the best-fit model maximizing the likelihood. To 
calculate dust lag, we fitted one optical (V band) and all NIR (J, H, 
$\mathrm{K_s}$ bands) light curves, simultaneously. The best-fit light curve and 
the probability distribution of the time delay are shown in Fig. \ref{fig:fig-12} 
for V band and NIR light curves. \textsc{javelin} lag distribution shows two peaks, 
a significant peak at $\sim$23 days and a small peak at $\sim$29 days. Considering the 
overall distribution, we calculated the lags and their uncertainties, which are 
given in Table \ref{tab:table-3} in the observer's frame. We do not find any 
wavelength-dependent lag from \textsc{javelin}. Almost similar lags between V 
and NIR bands are found from \textsc{javelin}.

\section{Discussion}
The lag obtained between the optical V band and the NIR $\mathrm{K_{s}}$ band is 
believed to
represent the inner edge of the dust torus as the temperature of the
dust at the wavelength of $\mathrm{K_{s}}$ band is close to the dust sublimation
temperature \citep{2017ApJ...843....3A}.  DRM observations through monitoring 
in the optical V band and NIR $\mathrm{K_{s}}$ band have enabled determination of the 
radius of the inner edge of the torus in about 40 AGN. A majority of those 
measurements are 
from the Multicolor Active Galactic Nuclei Monitoring (MAGNUM) project that 
includes 17 Seyfert 1 galaxies \citep{2014ApJ...788..159K}  and 22 
quasars \citep{2019ApJ...886..150M}  and a few objects are 
from other campaigns \citep{2014A&A...561L...8P,2015A&A...576A..73P,2018A&A...620A.137R,2018MNRAS.475.5330M}.

\subsection{Infrared lag and optical luminosity correlation}

\begin{figure}
\resizebox{9cm}{9cm}{\includegraphics{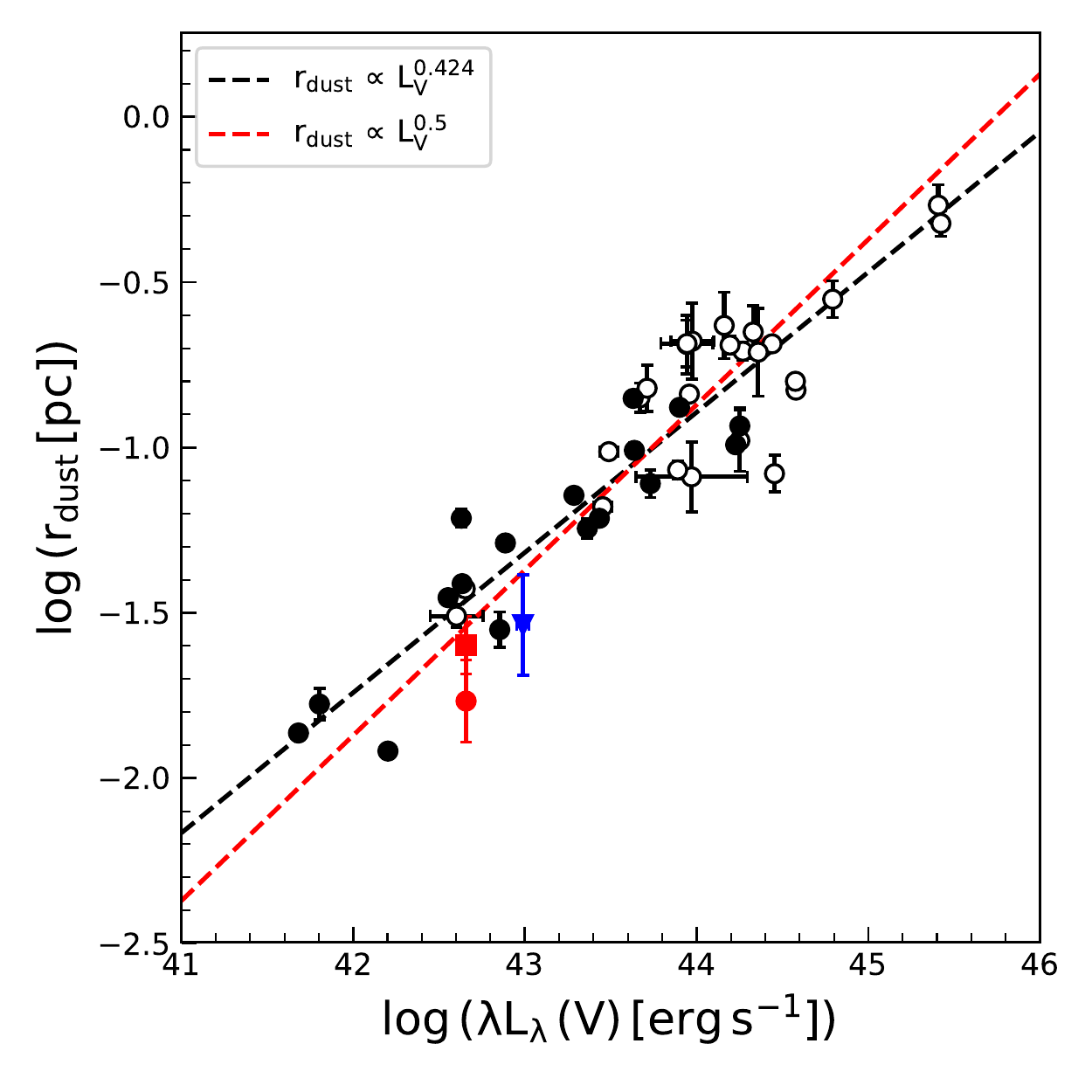}} 
\caption{Lag $-$ luminosity plot. The lags are in the rest frame of the sources. The empty and filled black circles represent the objects from \citet{2019ApJ...886..150M} and \citet{2014ApJ...788..159K}, respectively, triangular blue point represents H0507+164 obtained by \citet{2018MNRAS.475.5330M}, the red
circle and square points correspond to the lag of
Z229$-$15 for $\alpha$ obtained from observations and a constant $\alpha$ = 0.1, respectively, in the lag-luminosity plane.}
\label{fig:fig-13}
\end{figure}

According to \cite{1987ApJ...320..537B}, the size of the torus is expected to be
correlated with the luminosity of the accretion disk as 
$R_{\mathrm{torus}} \propto L^{0.5}$.  
We show in Fig. \ref{fig:fig-13} the plot of $R_{\mathrm{torus}}$ determined 
from K band  DRM lags 
and taken from literature 
\citep{2014ApJ...788..159K, 2018MNRAS.475.5330M, 2019ApJ...886..150M} against their optical luminosity.  In the same figure we show in 
dotted black line the linear regression relation of $R_{\mathrm{torus}} \propto L^{0.424}$ 
\citep{2019ApJ...886..150M} and $R_{\mathrm{torus}} \propto L^{0.5}$ \citep{2014ApJ...788..159K} by dotted red line. 
The source Z229$-$15 studied in this work for DRM is 
shown as a filled red circle in the Figure. Our lag measurement on the Z229$-$15 deviates from the linear regression line obtained by 
\cite{2019ApJ...886..150M}.  This deviation could be due to the adoption of
variable $\alpha$ in this work to correct for the contribution of the
AD to the observed NIR light curves. 
We  used $\alpha$ that was determined for each epoch using the near
simultaneous observations in B and V band, while for the lags obtained 
for the sources in Fig. \ref{fig:fig-13}, the authors have used a constant $\alpha$ 
to correct for the contribution of the AD to the NIR fluxes.
According to \cite{2008Natur.454..492K} the power law continuum from quasar
accretion disk can extend up to NIR with a shape characterised  as $F_{\nu} \propto
\nu^{1/3}$. AGN are known to show variations in the optical and NIR bands, but
the amplitude of flux variations need not be the same across wavelengths.
Thus, in addition to flux variations, AGN are also known to show spectral
variation \citep{2006ApJ...652L..13T, 2008Natur.454..492K}. Therefore, it might not be proper to assume a constant 
$\alpha$ from optical through NIR to correct for the IR 
contribution from the AD to the observed IR emission. Despite 
that, we redid the analysis by adopting a constant $\alpha$ of 0.1  as used by
\cite{2019ApJ...886..150M}. By this, we obtained a lag 
between V and $\mathrm{K_s}$ band of  30.04$^{+6.05}_{-4.82}$ days in the rest frame of the source as shown by the red square point which lies closer to the linear regression lines of \cite{2014ApJ...788..159K} and \cite{2019ApJ...886..150M} in Fig. \ref{fig:fig-13}. This 
lag of  30.04$^{+6.05}_{-4.82}$ days corresponds to a distance of  0.025 pc. 

From Fig. \ref{fig:fig-13},  $R_{\mathrm{torus}}$ is  found to be 
related to the luminosity with an index of  about 0.5 
(ie. $R_{\mathrm{torus}} \propto L^{0.5}$), however, there is scatter.
The scatter in the $R_{\mathrm{torus}} - L$ relation could come from 
various factors (cf. \citealt{2019ApJ...886..150M}), such as (a) assumption of 
a constant $\alpha$ on the removal of the AD component in the observed NIR 
light curves, (b) effect of viewing angle on the estimated lag 
\citep{2011ApJ...737..105K, 1992ApJ...400..502B}, (c) the effect of
accretion rate on the measured lag  and (d) the distribution of the dust 
and how it is illuminated by the central source \citep{2020ApJ...891...26A}. \cite{2014ApJ...788..159K}, by analysing
the residuals of the dust lag from the best fit linear regression could
not find systematic changes in the dust lag either with the
viewing angle or with the accretion rate. Reverberation studies
aimed in getting the size of BLR ($R_{\mathrm{BLR}}$) too found a linear correlation 
between the radius of the BLR and the luminosity with a power law index close 
to 0.5. However, recently from an analysis of the light curves obtained
for the SEAMBH program, 
\cite{2016ApJ...825..126D} found systematically lower BLR size for systems 
with high accretion rate. They found a slope $R_{\mathrm{BLR}} \propto L^{0.33}$, much
shallower than the slope of 0.533 found by \cite{2013ApJ...773...90G}. Given 
this observational finding, it is likely that accretion rate in an AGN will 
also have an influence on the derived dust reverberation lag. Homogeneous analysis on a larger  number of quasars are needed to arrive at a 
conclusion on the correlation between $R_{\mathrm{torus}}$ and accretion rate. 

\subsection{Size of the BLR and the dust torus}

\cite{2020MNRAS.491.4615K} by comparing the lag of the H${\beta}$ and the 
dust torus relative to the optical continuum for the  sources in
\cite{2014ApJ...788..159K} found that the dust lags are larger than the 
H${\beta}$ lag by a factor of about 4 ($R_{\mathrm{torus}} \sim 4 \times R_{\mathrm{BLR}}$).
According to  \citet{2014ApJ...788..159K} the dust lag is about five times 
larger than the BLR lag.  In Z229$-$15 the lag between the optical continuum 
and the H${\beta}$ emission line from spectroscopic reverberation 
observations is found to be $3.86^{+0.69}_{-0.90}$ days in the rest frame of 
the object \citep{2011ApJ...732..121B}.  For Z229$-$15, 
the dust lag obtained in this work is about a factor of 5.3  times larger than the BLR lag. This suggests that the BLR lies well inside the dust torus as
expected in the Unification model of AGN. There is a hint for wavelength 
dependent dust lags in our observations, however, due to the larger error bars in the lag between different wavelengths, the presence of wavelength dependent lag in Z229$-$15 cannot be unambiguously established. Sensitive, long term
and high cadence observations of Z229$-$15 are needed to confirm the presence of wavelength dependent lag if any.

\section{Summary}

We carried out DRM observations of the source Z229$-$15 in the optical (B, V) and NIR (J, H, $\mathrm{K_s}$) bands. We collected near simultaneous photometric data for a total of 48 epochs during the period 2017 July $-$ 2018 December. The results
of this work are summarized below:
\begin{enumerate}
\item The host galaxy is prominently seen in the observed B and V band images.
We devised a procedure to remove the contribution of the host galaxy to 
the observed optical light curves and get pure light curves of the optical
continuum source. 
\item The observed NIR light curves do have contribution from the accretion 
disk.  We removed this contribution by using the spectral index obtained between
B and V bands for each epoch. 
\item The host galaxy corrected optical light curves and the accretion disk component corrected NIR light curves show correlated variations among
themselves.
\item Using cross correlation function  analysis we found a rest-frame lag between 
 V and J band of $12.52^{+10.00}_{-9.55}$ days.
Similarly the lags between (V-H) and (V-K) are $15.63^{+5.05}_{-5.11}$ days and $20.36^{+5.82}_{-5.68}$ days, respectively. 
Due to the large error bars the presence of wavelength dependent lags if any 
could not be ascertained, instead the lag between different wavelengths are
consistent with each other.
\item Considering the lag between V and $\mathrm{K_s}$ to represent the inner edge of the
dust torus, we found that the inner edge of the dust torus in Z229$-$15 is 
at a distance of  0.017 pc from the central optical ionizing source.
\item The dust lag of 20.36 days is about a factor of 5.3 times larger than the BLR lag. This is similar to that known for other AGN.
\item Positioning our source in the $R_{\mathrm{torus}}-L$ relation known 
for a sample of about 40 DRM source, we found Z229$-$15 to lie below the 
$R_{\mathrm{torus}}-L$ found from DRM observations. However, using a 
constant spectral index of 0.1 to correct for the AD component in the NIR light curves, we found Z229$-$15 to move
closer to the known $R_{\mathrm{torus}}-L$ relation. We disfavour usage of constant $\alpha$ in DRM studies, as $\alpha$ is known to change with the brightness of the source.
\end{enumerate}

\section*{Acknowledgements}

 We thank the anonymous referee for her/his critical comments that helped to
improve our manuscript. We thank the staff at IAO, Hanle and CREST, Hoskote who helped with
the observations from the 2 m HCT. The facilities at IAO
and CREST are operated by the Indian Institute of Astrophysics, Bangalore.
The funding provided by the Humboldt Foundation, Germany is also thankfully acknowledged. AKM and RS thank the National Academy of Sciences, India (NASI), Prayagraj for funding  and Director, IIA for hosting and providing infrastructural support to this project.

\section*{Data availability}

The data underlying this article are available in the article and in its online supplementary material.


\bibliographystyle{mnras}

\bsp	
\label{lastpage}
\end{document}